\documentclass[fleqn,usenatbib,useAMS]{mnras}

\usepackage{graphicx}
\usepackage{enumerate}
\usepackage{newtxtext,newtxmath}
\usepackage{ae,aecompl}
\usepackage[T1]{fontenc}
\usepackage{gensymb}
\usepackage{mathtools}
\usepackage{physics}
\usepackage{natbib}
\usepackage{hyperref}
\usepackage{color}
\usepackage{ulem}
\usepackage{soul}
\usepackage{url}
\usepackage{xspace}
\usepackage{multirow}
\usepackage{xcolor}
\usepackage{array}
\usepackage[caption=false]{subfig}

\newcommand{\kms}{\,km\,s$^{-1}$} 
\newcommand{\mbh}{M_\mathrm{BH}}
\newcommand{\msun}{\mathrm{M}_\odot}

\title[WISDOM XXV.\ SMBH of NGC~1574]{WISDOM Project -- XXV. Improving the CO-dynamical supermassive black hole mass measurement in the galaxy NGC~1574 using high spatial resolution ALMA observations}

\author[H.\ Zhang et al.]{
Hengyue Zhang,$^{1}$\thanks{E-mail: hengyue.zhang@physics.ox.ac.uk}
Martin Bureau,$^{1}$\thanks{E-mail: martin.bureau@physics.ox.ac.uk}
Ilaria Ruffa,$^{2,3,4}$
Timothy A. Davis,$^{2}$
Pandora Dominiak,$^{1}$
\newauthor{
Jacob S. Elford,$^{5}$
Federico Lelli,$^{4}$
Thomas G. Williams$^{1}$
}
\\
$^{1}$Sub-department of Astrophysics, Department of Physics, University of Oxford, Denys Wilkinson Building, Keble Road, Oxford, OX1~3RH, UK \\
$^{2}$School of Physics \& Astronomy, Cardiff University, Queens Buildings, The Parade, Cardiff, CF24~3AA, UK \\
$^{3}$INAF - Istituto di Radioastronomia, via P.\ Gobetti 101, 40129 Bologna, Italy \\
$^{4}$INAF, Arcetri Astrophysical Observatory, Largo Enrico Fermi 5, I-50125 Florence, Italy\\
$^{5}$Instituto de Estudios Astrof\'{i}sicos, Facultad de Ingenier\'{i}a y Ciencias, Universidad Diego Portales, Av. Ej\'{e}rcito Libertador 441, Santiago, Chile\\
}

\pubyear{2025}

\begin{document}
\label{firstpage}
\pagerange{\pageref{firstpage}--\pageref{lastpage}}
\maketitle

\begin{abstract}
We present a molecular gas dynamical supermassive black hole (SMBH) mass measurement in the nearby barred lenticular galaxy NGC~1574, using Atacama Large Millimeter/sub-millimeter Array observations of the $^{12}$CO(2-1) emission line with synthesised beam full-widths at half-maximum of $0\farcs078\times0\farcs070$ ($\approx7.5\times6.7$~pc$^2$). The observations are the first to spatially resolve the SMBH's sphere of influence (SoI), resulting in an unambiguous detection of the Keplerian velocity increase due to the SMBH towards the centre of the gas disc. We also detect a previously known large-scale kinematic twist of the CO velocity map, due to a position angle (PA) warp and possible mild non-circular motions, and we resolve a PA warp within the central $0\farcs2\times0\farcs2$ of the galaxy, larger than that inferred from previous intermediate-resolution data. By forward modelling the data cube, we infer a SMBH mass of $(6.2\pm1.2)\times10^7$~M$_\odot$ ($1\sigma$ confidence interval), slightly smaller than but statistically consistent with the SMBH mass derived from the previous intermediate-resolution data that did not resolve the SoI, and slightly outside the $1\sigma$ scatter of the SMBH mass -- stellar velocity dispersion relation. Our measurement thus emphasises the importance of observations that spatially resolve the SMBH SoI for accurate SMBH mass measurements and gas dynamical modelling.
\end{abstract}



\section{Introduction}

Supermassive black holes (SMBHs) residing at the centres of nearly all massive galaxies (with total stellar masses $M_*\gtrsim 10^9$~$\msun$) play a pivotal role in galaxy evolution. This is implied by the tight scaling relations between black hole mass ($\mbh$) and several host galaxy properties, such as the stellar velocity dispersion within one effective radius
($\sigma_\mathrm{e}$; e.g.\ \citealp{Ferrarese_2000, Gebhardt_2000, Gultekin_2009}), bulge mass and total stellar mass \citep[e.g.][]{Beifiori_2012, Kormendy_2013}. These correlations suggest that SMBHs co-evolve with their host galaxies, growing through gas accretion from the interstellar media (ISM) while driving or regulating star formation by altering the cold gas contents of the galaxies (see e.g.\ \citealt{Kormendy_2013} and \citealt{D'Onofrio_2021} for reviews). However, the exact physical mechanism(s) responsible for this self-regulation are still under debate \citep[e.g.][]{Alexander_2012, Morganti_2017, Harrison_2018}, and the detailed history of this co-evolution across cosmic time remains a puzzle \citep[e.g.][]{Merloni_2010, Suh_2020, Pensabene_2020}. This is due to the remaining uncertainties in the SMBH -- galaxy property scaling relations at redshift $z=0$ and the difficulty of measuring such correlations at higher redshifts. It is thus crucial to obtain more direct and precise SMBH mass measurements in nearby galaxies, to better constrain the scaling relations in the local Universe, while providing calibrations for indirect high-redshift SMBH mass estimation methods such as reverberation mapping \citep[e.g.][]{Blandford_1982, Cackett_2021} and the single-epoch virial method \citep[e.g.][]{Vestergaard_2006, Maiolino_2024}.

Direct and reliable SMBH mass measurements come from spatially resolving and modelling the kinematics of matter (hereafter "kinematic tracer") within the SMBHs' spheres of influence (SoI), within which the SMBH masses dominate over the enclosed stellar masses. The most commonly used type of kinematic tracer is stars \citep[e.g.][]{Cappellari_2002b, Krajnovic_2009, Simon_2024}, as stellar dynamical modelling is in principle applicable to every galaxy. However, this method has been demonstrated to favour relatively dust-free early-type galaxies (ETGs; e.g.\ \citealp{Kormendy_2013}), as dust attenuates starlight and complicates measurements of stellar kinematics. 
Kinematics traced by ionised gas \citep[e.g.][]{Ferrarese_1996, Sarzi_2001, Walsh_2013} also suffers from dust extinction and is often influenced by non-gravitational forces such as turbulence, shocks and radiation pressure \citep[e.g.][]{Neumayer_2007, Osorno_2023}, making dynamical modelling difficult and potentially biased. Another frequently used kinematic tracer is megamasers, which lie very close to the SMBH and have provided the most precise SMBH mass measurements to date \citep[e.g.][]{Herrnstein_2005, Kuo_2011, Pesce_2020}. Nevertheless, megamaser emission originates almost exclusively from the nuclear activity of Seyfert 2 active galactic nuclei (AGN), so this method is heavily biased towards galaxies with $\sim10^7$~$\msun$ SMBHs \citep{Zhang_2024}. 

A novel method to measure SMBH masses, observing and modelling the kinematics of nuclear cold molecular gas discs, emerged in recent years thanks to the extraordinary sensitivity and angular resolution of the latest generation of sub-millimetre interferometers (see \citealt{Ruffa_2024} for a review). The approach is highly complementary to traditional methods and is applicable to a broad range of galaxy masses and morphological types, although modelling objects with significant cold gas non-circular motions (mostly late-type galaxies) can be more challenging \citep[e.g.][]{Combes_2019, Lang_2020}. Moreover, observations of molecular gas kinematics using the most extended configurations of the Atacama Large Millimeter/sub-millimeter Array (ALMA) can probe the circumnuclear discs down to the same physical scales (in the unit of the Schwarzschild radius, $R_\mathrm{Sch}\equiv2GM_\mathrm{BH}/c^2$, where $G$ is the gravitational constant and $c$ the speed of light) as observations of megamasers with very long baseline interferometry, in turn enabling some of the most precise SMBH mass measurements to date \citep{Zhang_2025}. The mm-Wave Interferometric Survey of Dark Object Masses (WISDOM) project has obtained a substantial number of high spatial resolution CO observations with (primarily) ALMA and has so far presented SMBH mass measurements of eleven massive ETGs \citep{Davis_2013b, Onishi_2017, Davis_2017, Davis_2018, Smith_2019, North_2019, Smith_2021, Ruffa_2023, Dominiak_2025_submitted, Zhang_2025}, a dwarf elliptical galaxy \citep{Davis_2020} and a peculiar luminous infrared galaxy with central spiral arms and a Seyfert nucleus \citep{Lelli_2022}. Other researchers have also published molecular gas dynamical SMBH mass measurements of thirteen additional ETGs \citep{Barth_2016, Boizelle_2019, Nagai_2019, Ruffa_2019b, Boizelle_2021, Cohn_2021, Kabasares_2022, Nguyen_2022, Cohn_2023, Cohn_2024, Dominiak_2024} and three barred spiral galaxies \citep{Onishi_2015, Nguyen_2020, Nguyen_2021}.

This paper presents a SMBH mass measurement in the galaxy NGC~1574 using ALMA observations of the $^{12}$CO(2-1) emission line with synthesised beam full-widths at half-maximum (FWHM) of $0\farcs078\times0\farcs070$ ($\approx7.5\times6.7$~pc$^2$; geometric mean $0\farcs074$ or $7.1$~pc). A previous CO-dynamical SMBH mass measurement in this galaxy, performed by \citet{Ruffa_2023}, inferred a mass of $(1.0\pm0.2)\times10^8$~$\msun$ ($1\sigma$ uncertainty here and throughout this paper) using ALMA observations of the same line at an intermediate angular resolution ($0\farcs202\times0\farcs138$). The intermediate-resolution data however slightly under-resolve the angular size of the SoI (see Section~\ref{subsec:SoI}).
The new study presented in this paper thus aims to improve the precision and accuracy of the existing $\mbh$ measurement, using data that fully spatially resolve the SoI. This study is the second of a series of high-angular resolution SMBH mass measurements performed by WISDOM, following the measurement of the SMBH mass of NGC~383 by \citet{Zhang_2025}. Works like these can reveal previously spatially-unresolved features of the circumnuclear molecular gas discs, such as small-scale warps and non-circular motions \citep[e.g.][]{Lelli_2022, Zhang_2025}, which could systematically bias SMBH mass measurements using molecular gas observations that only marginally resolve (or under-resolve) the SoI. High spatial resolution observations of molecular gas can also probe the circumnuclear disks of SMBHs down to a few times $10^4R_\mathrm{Schw}$, revealing the central fuelling and feedback mechanisms of SMBHs \citep[e.g.][]{Zhang_2025}.

This paper is structured as follows. Section~\ref{sec: observ} introduces the target, describes the ALMA observations and our data reduction procedures, and presents the final data product. Our molecular gas dynamical modelling technique and its results, including the best-fitting SMBH mass, are presented in Section~\ref{sec: modelling}. Section~\ref{sec: discuss} discusses the sources of statistical and systematic uncertainties. We conclude in Section~\ref{sec: conclude}.

\section{ALMA observations} \label{sec: observ}

\subsection{Target: NGC~1574} \label{subsec: target}

NGC~1574 is a relatively face-on ($i\approx27\degree$; \citealp{Ruffa_2023}) lenticular galaxy, one of the $46$ known members of the Dorado group \citep{Maia_1989}. It contains a nuclear stellar bar \citep[e.g.][]{Phillips_1996, Laurikainen_2011, Gao_2019}, which might be associated with the signs of weak non-circular motions detected in previous ALMA CO observations \citep{Ruffa_2023}. The luminosity distance of NGC~1574 is $19.9\pm2.0$~Mpc, derived from the surface brightness fluctuation (SBF) distance modulus reported by \citet{Tonry_2001}. Assuming a cold dark matter cosmology with a Hubble constant $H_0=70$~\kms~Mpc$^{-1}$, a matter density (relative to the critical density) $\Omega_\mathrm{M}=0.3$ and a cosmological constant (relative to the critical density) $\Omega_{\Lambda}=0.7$ today, the SBF distance corresponds to a redshift $z=0.0046$ and an angular-diameter distance of $19.7\pm2.0$~Mpc. At this distance, an angle of $1^{\prime\prime}$ corresponds to a physical separation of $\approx 95$~pc. This distance is consistent with that adopted in the dynamical modelling of \citeauthor{Ruffa_2023} (\citeyear{Ruffa_2023}; 19.76~Mpc). We have confirmed that the angular-diameter distance of $19.3$~Mpc reported by \citet{Ruffa_2023} is a typographical error.

Assuming a SMBH mass $M_\mathrm{BH}=(1.0\pm 0.2)\times10^8$~$\msun$ (\citealt{Ruffa_2023}) and $\sigma_\mathrm{e}=216\pm16$~\kms\ \citep{Bernardi_2002}, the radius of the SMBH SoI ($R_\mathrm{SoI}$) of NGC~1574, as estimated using $R_\mathrm{SoI}\approx GM_\mathrm{BH}/\sigma_\mathrm{e}^2$, is $9.2$~pc or $0\farcs1$.
Our ALMA observations (see Section~\ref{subsec: data}) with a mean synthesised beam FWHM of $0\farcs074$ ($7.1$~pc) should thus be the first to spatially resolve the SMBH SoI of NGC~1574, enabling for the first time a highly reliable SMBH mass determination in this object. We provide a more accurate estimate of the size of the SoI of NGC~1574 in Section~\ref{subsec:SoI}, using our updated SMBH mass and a more formal SoI definition.

\subsection{Observations and data reduction}
\label{subsec: data}

The $^{12}$CO(2-1) emission line of NGC~1574 was observed with the ALMA 12-m array in band 6 using configuration C-8 on 2023 September 16 as part of project 2022.1.01122.S (PI: M.\ Bureau). The observations consist of two tracks with baseline ranges of $92$~m to $8.8$~km and $83$~m to $8.5$~km and on-source observing times of $5977$ and $5984$~s, respectively. We also use three prior intermediate-resolution ALMA observations of the same line to improve the $uv$-plane coverage. Two tracks (both part of project 2015.1.00419.S, PI: T.\ Davis) used the 12-m array and have baseline ranges of $15$~m to $0.7$~km and $80$~m to $3.2$~km and on-source times of $393$ and $756$~s, respectively. The other track (project 2016.2.00053.S, PI: L.\ Liu) used the 7-m Atacama Compact Array (ACA) and has baselines ranging from $9$ to $49$~m and a total on-source time of $2359$~s. These three prior tracks were presented by \citet{Ruffa_2023}, who used them to derive the previous SMBH mass. The properties of all the tracks used in this study are summarised in Table~\ref{tab:Observing Tracks}.

\begin{table*}
    \renewcommand\thetable{1}
	\centering
	\caption{Properties of the ALMA observing tracks analysed in this paper.}
	\label{tab:Observing Tracks}
	\begin{tabular}{llllcccll}
		\hline
		Project code & Date & Array & Baseline range & $T_\mathrm{obs}$ & MRS & FoV & Calibration & Reference\\
            & & & & (s) & (arcsec, kpc) & (arcsec, kpc)\\
            (1) & (2) & (3) & (4) & (5) & (6) & (7) & (8) & (9)\\ 
		\hline
		2015.1.00419.S & 2016 June 12 & 12-m & 15~m -- $0.7$~km & $\phantom{0}393$ & $\phantom{0}5.5$, $0.5$ & $24.4$, $2.3$ & Pipeline & \citet{Ruffa_2023} \\
  
		2015.1.00419.S & 2016 September 27 & 12-m & $80$~m -- $3.2$~km & $\phantom{0}756$ & $\phantom{0}3.3$, $0.3$ & $24.4$, $2.3$ & Pipeline & \citet{Ruffa_2023}\\
  
		2016.2.00053.S & 2017 July 24 & \phantom{0}7-m & $\phantom{0}9$~m -- $49$~m & $2359$ & $30.2$, $2.9$ & $41.8$, $4.0$ & Pipeline & \citet{Ruffa_2023}\\

        2022.1.01122.S & 2023 September 16 & 12-m & $92$~m -- $8.8$~km & $5977$ & $\phantom{0}1.0$, $0.1$ & $26.3$, $2.5$ & Pipeline & This work\\

        2022.1.01122.S & 2023 September 16 & 12-m & $83$~m -- $8.5$~km & $5984$ & $\phantom{0}1.0$, $0.1$ & $26.3$, $2.5$ & Pipeline & This work\\
		\hline
	\end{tabular} \\
        {{\sl Notes.} Columns: (1) Project code. (2) Observation date. (3) ALMA array. (4) Minimum and maximum baseline lengths. (5) Total on-source observing time ($T_\mathrm{obs}$). (6) Maximum recoverable scale (MRS), i.e.\ largest angular scale that can be recovered with the given array configuration. (7) Field of view (FoV), i.e.\ primary beam FWHM. (8) Calibration method. (9) Reference in which the track data were first presented.}
\end{table*}

Each of the five observing tracks has four spectral windows (SPWs). One is always centred at $\approx229.8$~GHz to target the redshifted $^{12}$CO(2-1) line (rest frequency $230.538$~GHz), covering a bandwidth of $1.875$~GHz ($\approx2400$~\kms) with $1920$ channels of $976.6$~kHz ($\approx1.3$~\kms) for the 12-m array observations and a bandwidth of $2$~GHz ($\approx2600$~\kms) with $2048$ channels of $976.6$~kHz ($\approx1.3$~\kms) for the ACA observations. In all observations, the other three SPWs are deployed to map continuum emission. In the two most recent, high angular resolution observing tracks, these three SPWs are centred at $212.8$, $214.8$ and $227.8$~GHz and have the same channel width and total bandwidth as the line SPW. The continuum SPWs of the three older observing tracks are centred at $231.7$, $245.2$ and $247.2$~GHz and have $128$ channels of $15.63$~MHz ($\approx20$~\kms), covering a total bandwidth of $2$~GHz.

For all the tracks, a standard calibration strategy was adopted, using a single bright quasar as both flux and bandpass calibrator and a second bright quasar as phase calibrator. The flux and bandpass calibrator of the high-resolution observing tracks is J0519-4546 and the phase calibrator is J0425-5331.

The raw data were calibrated with the \textsc{Common Astronomy Software Application} (\textsc{CASA}) package\footnote{\url{https://casa.nrao.edu/}} \citep{McMullin_2007}, version~6.4.1.12, using the standard script provided by the ALMA Science Archive\footnote{\url{https://almascience.eso.org/aq/}}.
We combined all the calibrated observing tracks using the \textsc{CASA} task {\tt concat}. The combined dataset was then imaged with \textsc{CASA} version~6.5.2 using the \texttt{tclean} task with a H{\"o}gbom deconvolver \citep{Hogbom_1974}. For both the line and continuum data, we chose a pixel size of $0\farcs02$ and an image size of $200 \times 200$ pixels, sufficient to properly sample the synthesised beam and cover the full spatial extent of the emission. Further details of the line data cube and the continuum image are presented below.

\subsection{Line emission}

\subsubsection{Imaging}

We separated the $^{12}$CO(2-1) line emission from the continuum emission in the $uv$-plane using the \textsc{CASA} task {\tt uvcontsub}, which fits a continuum model that is linearly-varying with frequency to interactively-identified line-free channels and then subtracts the best-fitting model from the calibrated visibilities. We then imaged the continuum-subtracted visibilities in the cube mode of \texttt{tclean} using Briggs weighting with a robust parameter of $1.0$. After some trials, we chose this weighting to ensure that the SMBH SoI is sufficiently spatially resolved while compensating for the data's relatively low signal-to-noise ratio ($S/N$). We adopted a channel width of $20$~\kms, sufficient for a reliable analysis of the gas kinematics \citep[e.g.][]{Ruffa_2023, Zhang_2025} while also achieving a reasonable $S/N$.
The channel velocities were computed in the rest frame, with the rest frequency of the $^{12}$CO(2-1) line ($230.538$~GHz) corresponding to a velocity of zero. The dirty (uncleaned) data cube was cleaned in interactively-identified regions of line emission down to a threshold of $1.5$~$\sigma_\mathrm{RMS}$, where $\sigma_\mathrm{RMS}$ is the root-mean-square (RMS) noise evaluated in line-free regions of the dirty cube. The final cleaned data cube has $\sigma_\mathrm{RMS}=0.15$~mJy~beam$^{-1}$ and synthesised beam FWHM of $0\farcs078\times0\farcs070$ ($\approx7.5\times6.7$~pc$^2$), with a position angle (PA) of $31\fdg5$. The data cube's average $S/N$ in regions of line emission is $\approx 4$.

\subsubsection{Moment maps}
\label{subsec: moments}

\begin{figure*}
\captionsetup[subfigure]{labelformat=empty}
    \centering
    \subfloat[]{
    \includegraphics[width=0.465\linewidth]{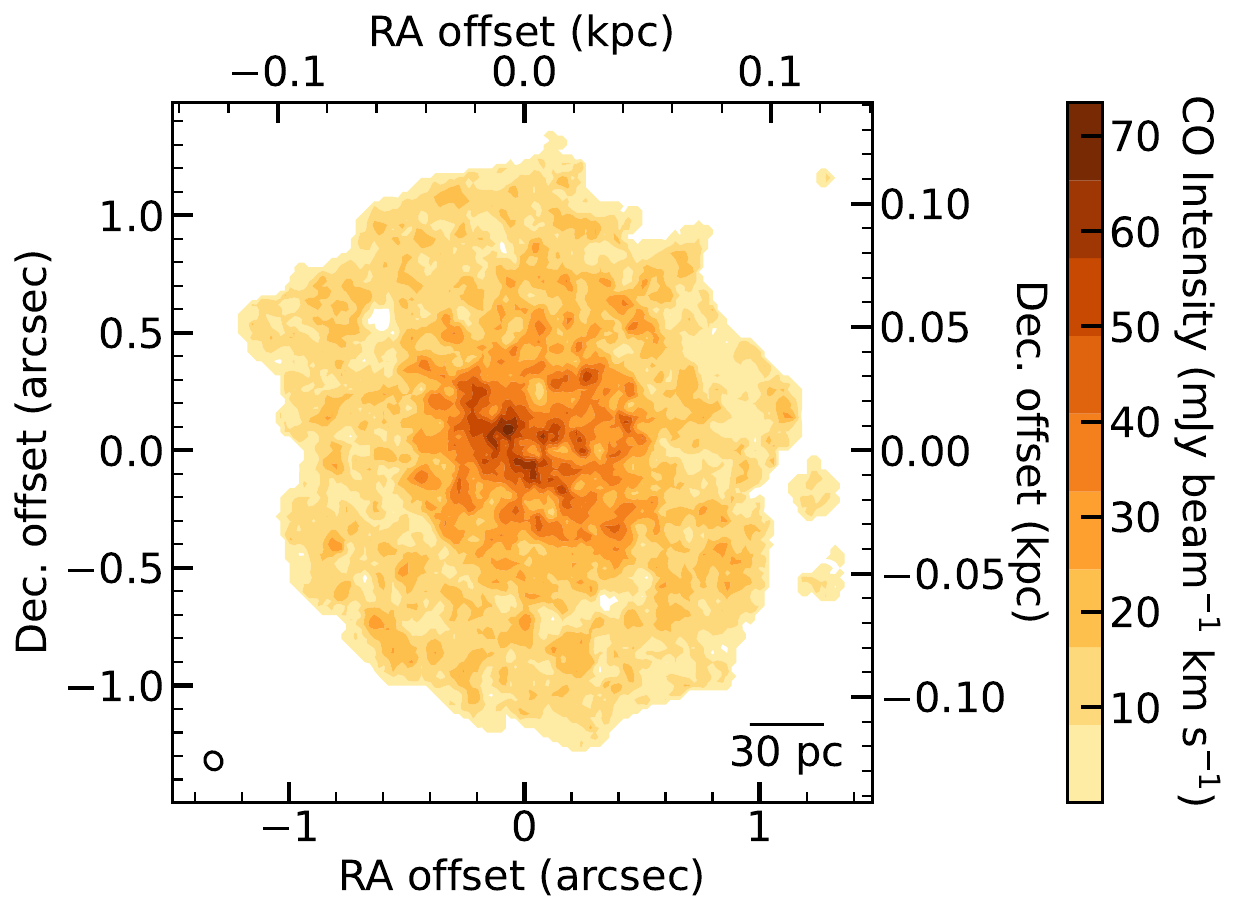}
    }
    \subfloat[]{
    \includegraphics[width=0.488\linewidth]{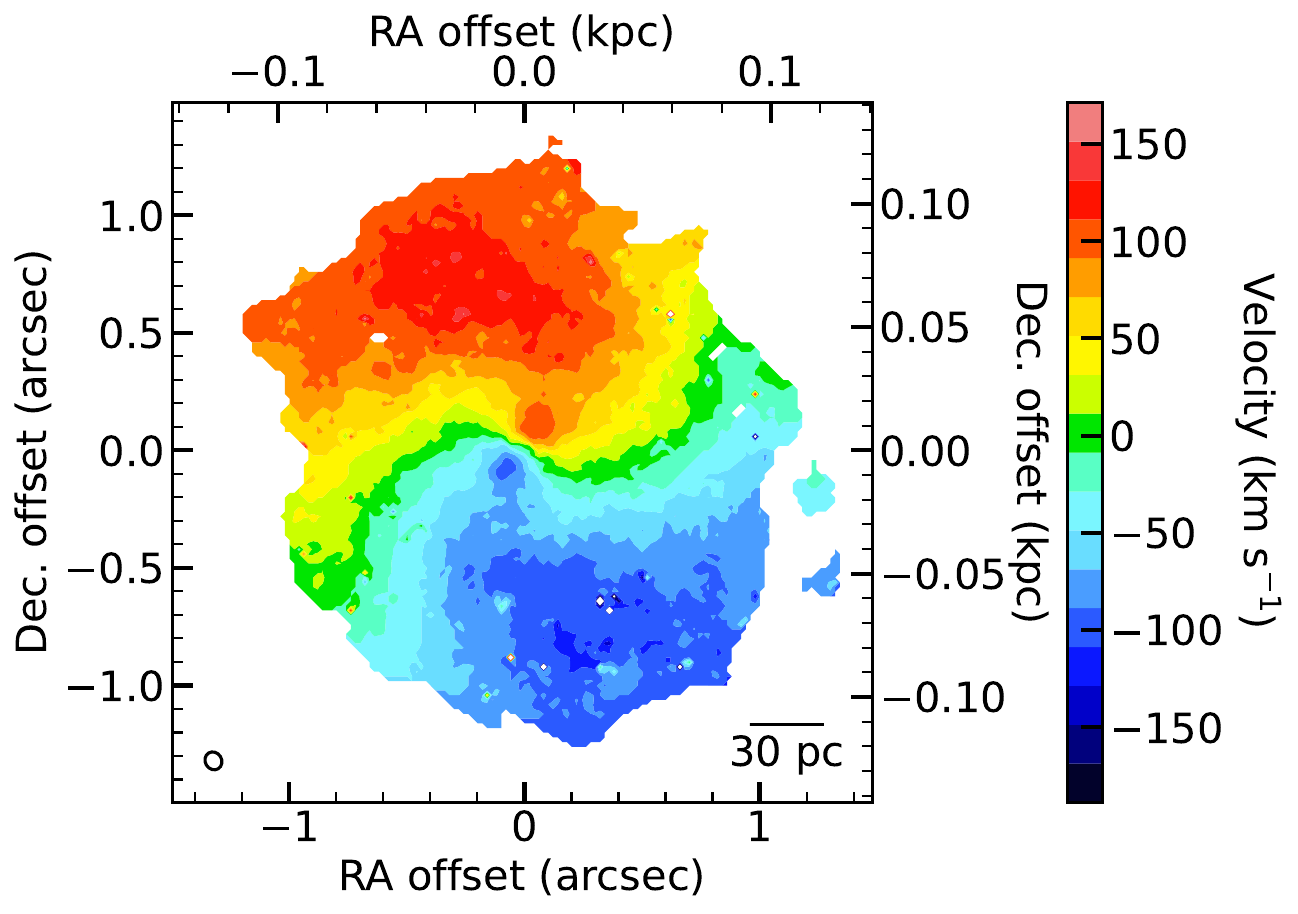}
    } \\
    \subfloat[]{\includegraphics[width=0.475\linewidth]{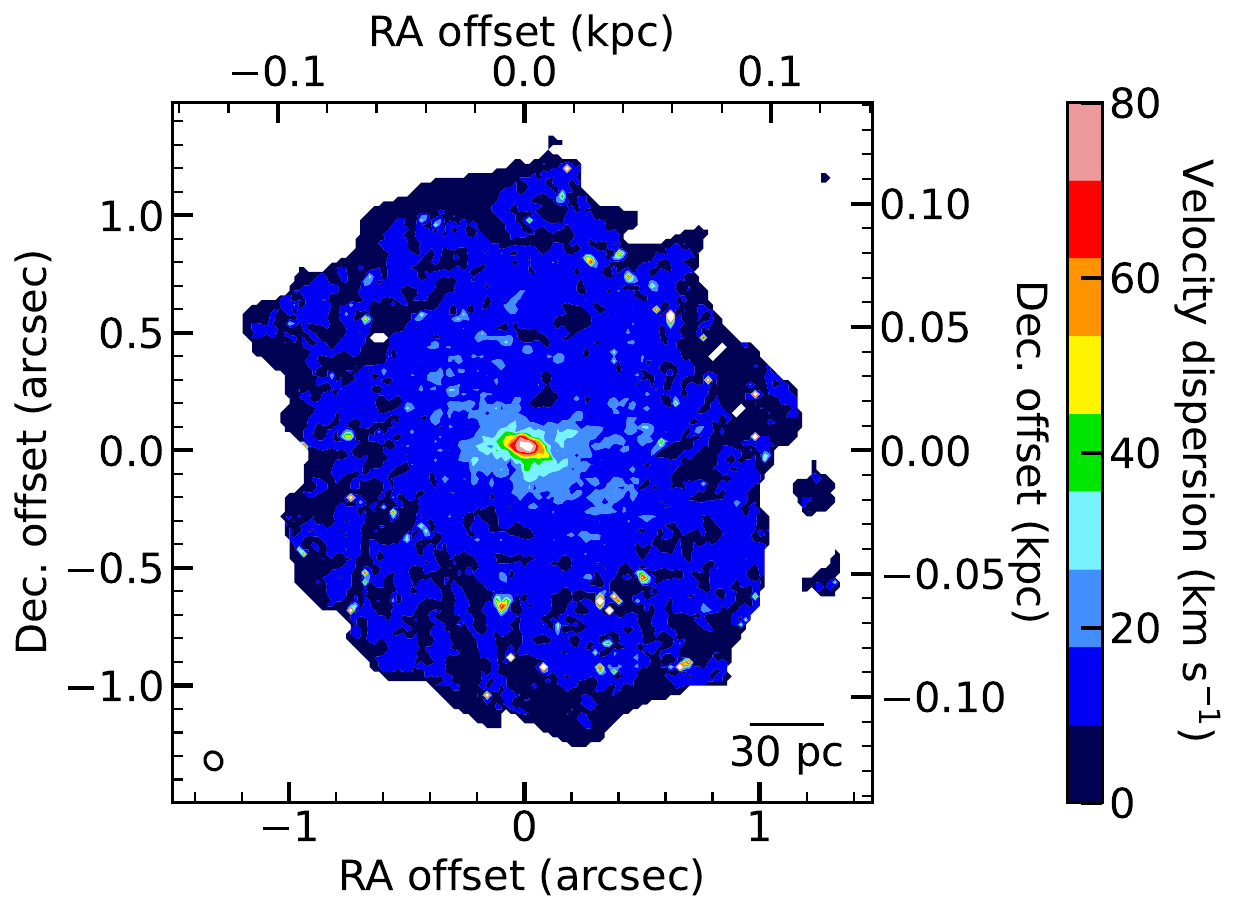}
    }
    \subfloat[]{\includegraphics[width=0.521\linewidth]{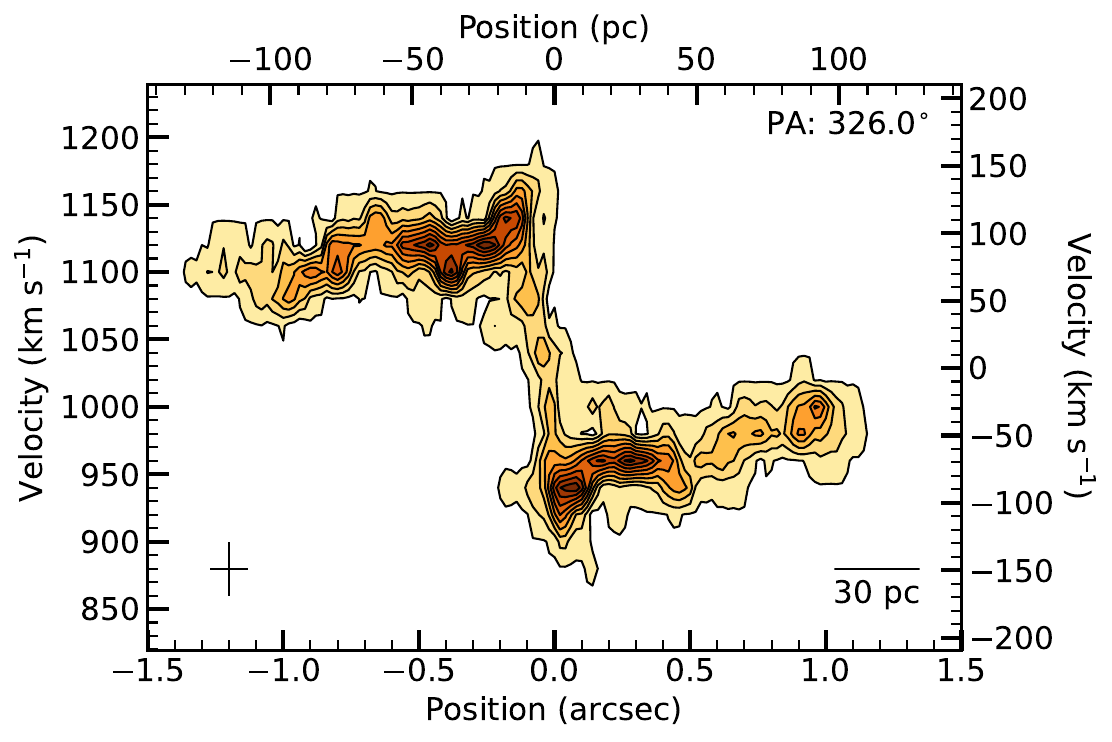}}
    \caption{Data products of NGC~1574 created from our ALMA $^{12}$CO(2-1) data cube. \textbf{Top-left:} Zeroth-moment (integrated-intensity) map. \textbf{Top-right:} First-moment (intensity-weighted mean line-of-sight velocity) map. \textbf{Bottom-left:} Second-moment (intensity-weighted line-of-sight velocity dispersion) map. The synthesised beam ($0\farcs078\times0\farcs069$) is shown as a black open ellipse in the bottom-left corner of each map. \textbf{Bottom-right:} Kinematic major-axis position-velocity diagram (PVD), covering the central $1\farcs5$ on either side of the kinematic centre. 
    The cross in the bottom-left corner shows the synthesised beam FWHM along the kinematic major axis and the channel width. A $30$~pc scale bar is shown in the bottom-right corner of each panel. In all panels, positions are measured relative to the best-fitting kinematic centre (see Section~\ref{subsec: results}); velocities relative to the best-fitting systemic velocity along the right axis of the PVD.
    }
    \label{fig:moments}
\end{figure*}

We adopt a masked-moment technique \citep{Dame_2011} to create the zeroth (integrated-intensity), first (intensity-weighted mean line-of-sight velocity) and second (intensity-weighted line-of-sight velocity dispersion) moment maps of the final data cube. To generate the mask, we take a copy of the data cube and smooth it spatially with a uniform filter of size $3$ times the mean synthesised beam FWHM, and spectrally with a Hanning window of $5$ channels. The mask includes all pixels of the smoothed data cube with a flux density larger than $1$~$\sigma_\mathrm{RMS}$ of the unsmoothed data cube. The moment maps are then created using only the masked pixels of the unsmoothed data cube and are shown in Figure \ref{fig:moments}. The filter sizes and the flux threshold were chosen by trial and error to minimise the noise in the moment maps while recovering most of the emission. We also note that the moment maps are generated only to help visualise the data cube. Our dynamical modelling described in Section~\ref{sec: modelling} uses the entire unmasked data cube rather than the moment maps.

The integrated-intensity map shows a molecular gas disc extending $\approx 2\farcs5 \times 2\farcs2$ ($\approx240\times210$~pc$^2$) in projection. The brightest CO emission is slightly offset from the geometric centre of the gas disc, by $\approx-0\farcs07$ ($\approx-7$~pc) in right ascension and $\approx0\farcs09$ ($\approx9$~pc) in declination, consistent with the offsets observed in \citet{Ruffa_2023} but smaller than the synthesised beam size of the old observations. Compared to the intermediate-resolution data cube presented by \citet{Ruffa_2023}, our high-resolution data cube reveals more molecular gas clumps, as it spatially resolves smaller physical scales ($\approx7$~pc). Analysis of these molecular gas clumps is however beyond the scope of this study.

The mean line-of-sight velocity map reveals a rotating disc with a prominent twist of the isovelocity contours across the entire gas distribution. Dynamical modelling of the intermediate-resolution data cube suggests that this twist is due to a linearly-varying PA warp, from $32\degree$ (measured counterclockwise from the celestial North to the redshifted side of the kinematic major axis) at the outer edge of the disc to $342\degree$ at its centre (after going through $0\degree$). Our higher-resolution data however indicate that the PA warp in the nuclear region is slightly larger than previously inferred (see Section~\ref{subsec: results}). Crucially, the velocity map also clearly reveals an increase (from the outside-in) of the mean line-of-light velocities within the central $\approx0\farcs15$ ($\approx15$~pc) in radius, consistent with the expected gas motion around a central SMBH and absent from previous observations.

The intensity-weighted line-of-sight velocity dispersion map reveals a radial gradient of the CO velocity dispersion ($\sigma_\mathrm{gas}$), the disc being mostly dynamically cold ($\sigma_\mathrm{gas}\leq20$~\kms) but with a velocity dispersion enhancement in its central regions ($\sigma_\mathrm{gas}$ up to $\approx80$~\kms). It is however likely that this central enhancement is primarily caused by observational effects such as beam smearing, which often substantially increase the observed velocity dispersions in areas with large velocity gradients (such as the central regions of galaxies). The observed enhancement therefore does not necessarily imply an intrinsic velocity dispersion gradient.

\subsubsection{Total flux}

\begin{figure}
    \centering
\includegraphics[width=\linewidth]{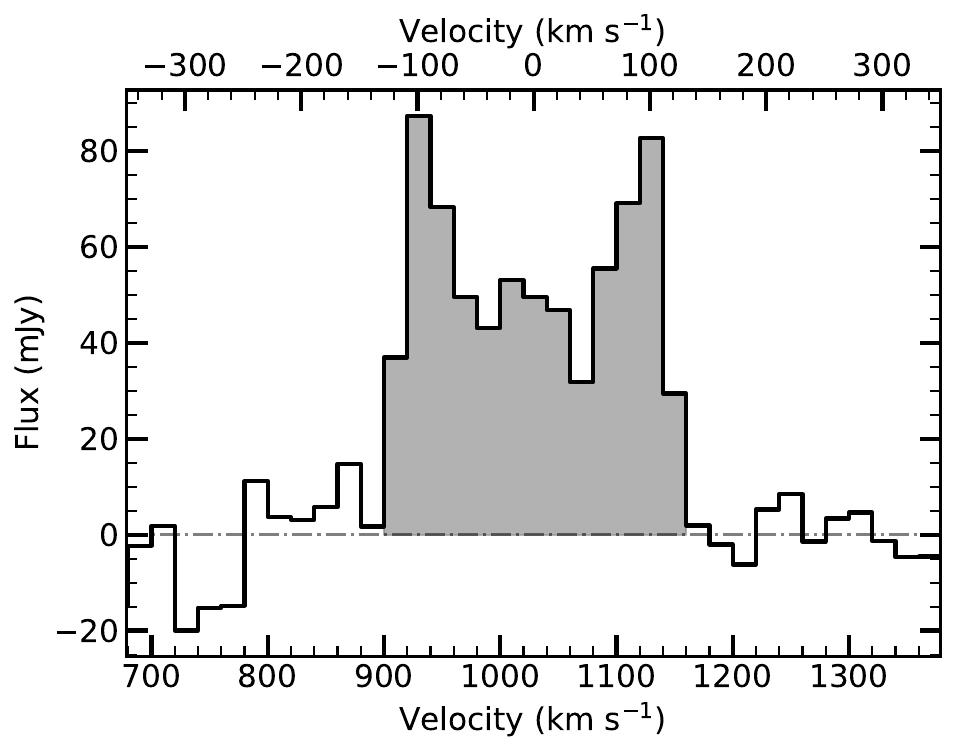}
    \caption{Integrated $^{12}$CO(2-1) spectrum of NGC~1574, extracted from the central $2\farcs6 \times 2\farcs6$ region of the data cube, covering all of the detected emission. Velocities are measured relative to the best-fitting systemic velocity along the top axis (see Section~\ref{subsec: results}). The dot-dashed line indicates the zero flux level. The shaded region highlights the channels within the line FWZI, used to measure the total flux. The spectrum shows the typical double-horned shape of a rotating disc.
    }
    \label{fig:spec}
\end{figure}

We obtain the spatially-integrated $^{12}$CO(2-1) spectrum of NGC~1574 by summing all pixels (without applying any masks) within the central $2\farcs6\times2\farcs6$ region of the data cube, that covers all of the detected emission. The integrated spectrum (Figure~\ref{fig:spec}) exhibits the typical double-horned shape of a rotating disc and only mild asymmetries. To compute the total $^{12}$CO(2-1) flux of the data cube, we then integrate this spectrum across its full-width at zero-intensity (FWZI; the velocity range including all channels with line emission above $3\sigma$). The total flux is $13.8\pm 0.1\pm1.4$~Jy~\kms, where the second error is systematic and results from the typical $\approx10\%$ ALMA flux calibration uncertainty.

We note, however, that the total flux measured is sensitive to the cleaning depth. When we create another data cube with the same imaging procedure but a shallower cleaning depth of $2.0$~$\sigma_\mathrm{RMS}$, the resulting total flux is $15.6\pm 0.1\pm1.6$~Jy~\kms. This is because fitting a clean Gaussian beam to the dirty beam systematically underestimates the effective area of the synthesised beam, especially for multi-configuration data, causing an overestimate of the total flux for any finite cleaning depth (see e.g.\ \citealt{Jorsater_1995}, \citealt{Bureau_2002} and \citealt{Zhang_2025} for a detailed explanation of the issue). We thus follow the prescription of \citet{Jorsater_1995} to derive the true total flux (corresponding to an infinite cleaning depth) from the measured cleaned fluxes and total fluxes of the two data cubes cleaned to different depths. The corrected total flux obtained with this method is $10.1\pm0.1\pm1.0$~Jy~\kms. This is smaller than the integrated flux of $17.9\pm0.1\pm 1.8$~Jy~\kms\ computed from the intermediate-resolution data cube of \citet{Ruffa_2023}, but it agrees perfectly with the corrected total flux of these intermediate-resolution data (also $10.1\pm 0.1\pm1.0$~Jy~\kms) after applying the same flux correction procedure. We thus adopt this corrected total flux, but stress that the total flux and associated flux correction have no impact on our derived SMBH mass. The measured FWZI ($\approx260$~\kms) and FWHM ($\approx220$~\kms) of the $^{12}$CO(2-1) line both agree with those reported by \citet{Ruffa_2023}. 

\subsubsection{PVD}

The bottom-right panel of Figure~\ref{fig:moments} shows the kinematic major-axis position-velocity diagram (PVD) of NGC~1574 extracted along a PA of $326\fdg0$ (the best-fitting innermost kinematic PA derived in Section~\ref{subsec: results}) by summing all the flux within a $4$-pixel-wide (approximately the synthesised beam FWHM) pseudo-slit centred on the major axis. When creating the PVD, we applied a mask similar to the one used for the moment maps, except that the data cube was spatially smoothed using a Gaussian filter with a FWHM equal to $1.5$ times that of the synthesised beam. We then selected all the pixels of the smoothed data cube with a flux density larger than $0.7$~$\sigma_\mathrm{RMS}$ of the unsmoothed data cube. This approach avoids over-smoothing the central region, which could result in a mask that excludes faint emission in central high-velocity channels. The filter sizes and the flux threshold
were again chosen by trial and error to ensure that the kinematics closest to the SMBH is clearly revealed while minimising noise. The resultant PVD reveals a prominent increase of the (projected) line-of-sight velocities toward the centre, up to $\approx170$~\kms, similar to the Keplerian circular velocity curve expected in the presence of a SMBH. By comparison, the intermediate-resolution observations of \citet{Ruffa_2023} only revealed mild evidence of a central velocity enhancement, reaching a (projected) line-of-sight velocity of only $\approx120$~\kms\ in the innermost regions due to a resolution insufficient to detect higher-velocity gas closer to the SMBH (beam smearing effects). This clearly demonstrates that the higher-resolution ALMA CO observations presented in this work are the first to spatially resolve the SMBH SoI of NGC~1574, in turn enabling a more accurate SMBH mass measurement, which we present in the following section.

\subsection{Continuum emission}

\begin{figure}
    \centering
    \includegraphics[width=\linewidth]{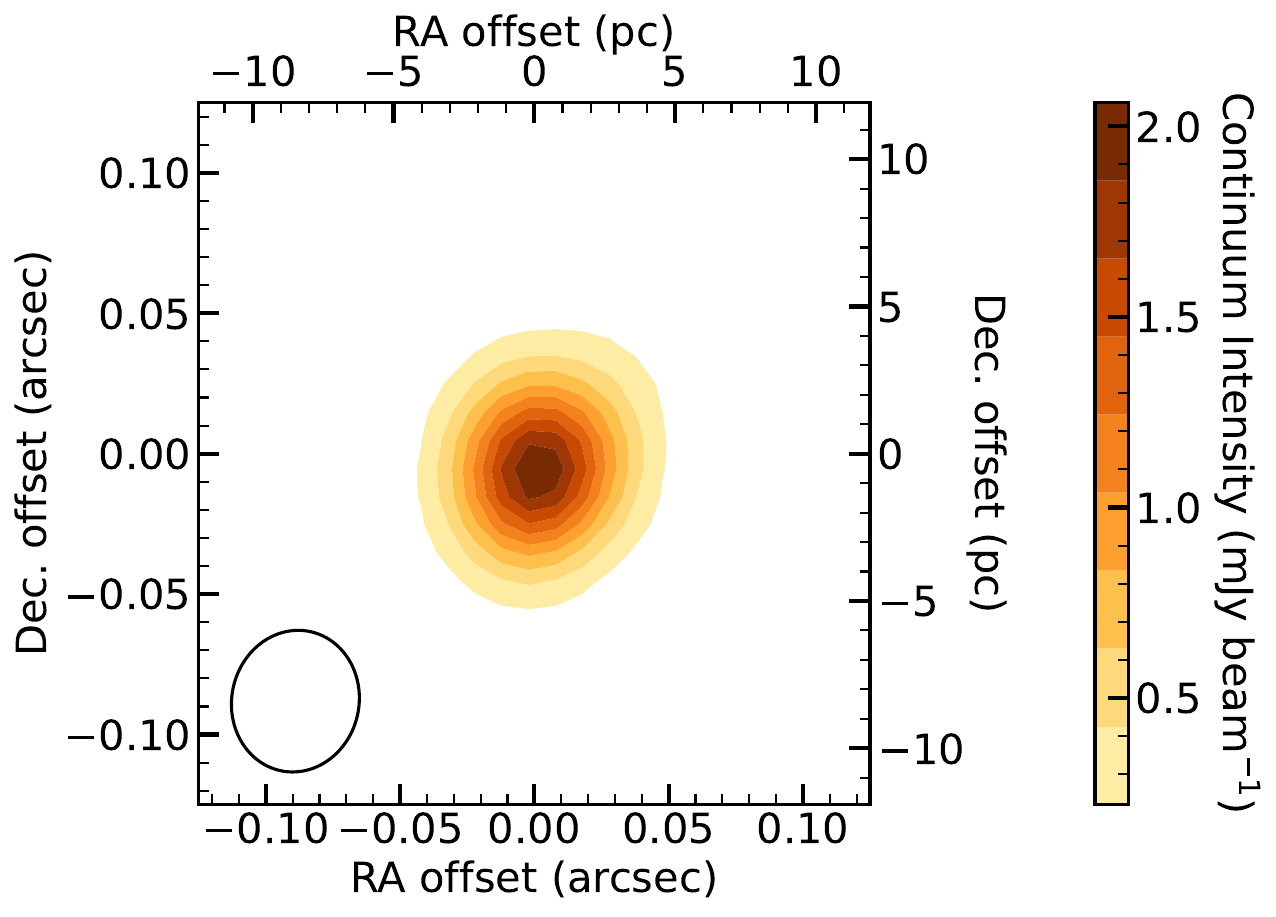}
    \caption{Central region ($0\farcs25\times0\farcs25$) of the NGC~1574 $1.3$-mm continuum image, showing the only source detected. Contour levels are equally spaced between the peak intensity of $2.02\pm0.02$~mJy~beam$^{-1}$ and $20$ times the RMS noise. The synthesised beam ($0\farcs078\times0\farcs069$) is shown in the bottom-left corner as a black open ellipse. The source is only marginally spatially resolved.}
    \label{fig:continuum}
\end{figure}

A continuum image was created using the multi-frequency synthesis mode of the \textsc{CASA} task \texttt{tclean} and Briggs weighting with a robust parameter of $0.5$. We choose this weighting to marginally spatially resolve the source while still reaching a $S/N$ of $\approx200$. This results in synthesised beam FWHM of $0\farcs051\times0\farcs047$ ($\approx4.9\times4.5$~pc$^2$) at a PA of $-18\fdg8$ and a RMS noise of $11$~$\mu$Jy~beam$^{-1}$. The continuum image, shown in Figure~\ref{fig:continuum}, reveals only a point-like source spatially coincident with the galaxy's kinematic centre (the best-fitting SMBH position in Section \ref{subsec: results}). We fit a two-dimensional (2D) Gaussian to the source with the \textsc{CASA} task \texttt{imfit}, revealing a marginally-resolved source with a deconvolved FWHM size of $0\farcs024\times0\farcs016$ ($\approx2.3\times1.5$~pc$^2$) and an integrated flux density of $2.37\pm0.03\pm0.24$~mJy (where the second uncertainty is systematic) at a central frequency of $231$~GHz.

\section{Dynamical modelling}
\label{sec: modelling}

Our procedure to model the gas dynamics and infer the SMBH mass in NGC~1574 is similar to those used in previous WISDOM SMBH papers \citep[e.g.][]{Davis_2017, Davis_2018, Smith_2019, Smith_2021, Zhang_2025}. In this section, we thus provide only an outline of our method and focus on the specifics of modelling NGC~1574.

We forward model the observed gas kinematics using the publicly-available \textsc{Python} version of the \textsc{Kinematic Molecular Simulation}\footnote{\url{https://github.com/TimothyADavis/KinMS_fitter}\\
\phantom{xx}\url{https://kinms.space/}} (\textsc{KinMS}; \citealt{Davis_2013a}) tool. \textsc{KinMS} takes an input model of a galaxy's gas distribution and circular velocity profile to construct a simulated data cube, taking into account observational effects such as beam smearing, spatial and velocity binning, and line-of-sight projection. Then, the simulated data cube is fitted to the observed data cube using \textsc{GAStimator}\footnote{\url{https://github.com/TimothyADavis/GAStimator}}, a Markov chain Monte Carlo (MCMC) Gibbs sampler, to infer the best-fitting model parameters and their posterior distributions. We define the $\chi^2$ to be
\begin{equation}
    \chi^2\equiv\frac{1}{\sigma_\mathrm{RMS}^2}\, \sum_{i} (\mathrm{data}_i-\mathrm{model}_i)^2\,\,\,,
\end{equation}
where the summation is over all pixels of the data cube. We further re-scale the $\chi^2$ by a factor of $1/\sqrt{2N}$, where $N=36376$ is the number of pixels with detected emission, defined as the pixels included in the mask of Section~\ref{subsec: moments}. The $\chi^2$ rescaling is equivalent to inflating the uncertainties by a factor of $(2N)^{1/4}$. This approach provides a realistic estimate of the systematic uncertainties, that usually dominate over the statistical uncertainties in large data sets. It was first proposed by \citet{vdBosch_2009} when using $\chi^2$ confidence levels and was later adapted by \citet{Mitzkus_2017} for Bayesian methods. \citet{Smith_2019} and \citet{Zhang_2025} showed that the uncertainties of the gas dynamical parameters estimated using this approach are consistent with those evaluated from bootstrapping and are thus reliable. 

We also assume a diagonal covariance matrix when evaluating the $\chi^2$, ignoring the spatial correlations between nearby pixels introduced by the synthesised beam, as inverting the full covariance matrix would be computationally intractable. \citet{Davis_2017} and \citet{Davis_2018} adopted the full covariance matrix, but had to limit the fitting to a small region to reduce the number of pixels and thus compensate for the substantial increase of the computational time. As our high-resolution data cube has a huge number of pixels, limiting the number of pixels in the fitting would bias the results much more than assuming a diagonal covariance matrix. In any case, the covariance is much smaller than the $\chi^2$ rescaling described above, and so would have a negligible impact on the best-fitting SMBH mass and its uncertainty.

\subsection{Gas distribution}

Although our high spatial resolution observations spatially resolve clumpy structures in the intensity distribution (see Figure~\ref{fig:moments}), the azimuthally-averaged intensity profile is still consistent with an exponential profile, so we adopt the approach of \citet{Ruffa_2023} and model the spatial distribution of the CO(2-1) emission as a simple exponential disc:
\begin{equation}
    \Sigma_\mathrm{gas}(R) \propto e^{-R/R_\mathrm{s}}\,\,\,,
\end{equation}
where $\Sigma_\mathrm{gas}$ is the surface brightness of the CO gas, $R$ the galactocentric cylindrical radius and $R_\mathrm{s}$ the exponential disc scale length, a free parameter of the model. We have confirmed that adopting a non-parametric gas distribution, such as using the CLEAN components generated by \texttt{tclean} as the input cloud model \citep[e.g.][]{Smith_2019,Zhang_2025}, does not affect the best-fitting parameters presented in Section~\ref{subsec: results} by more than $1\sigma$ but introduces additional systematic uncertainties due to the finite cleaning depth. We have also confirmed that the best-fitting exponential surface brightness profile agrees well with the observed deconvolved, azimuthally-averaged gas intensity profile obtained by sampling the CLEAN components.

\subsection{Mass model}

To model the mass distribution of NGC~1574, we assume that the total mass in the central $\approx 0.3\times0.3$~kpc$^{2}$ region we observe is dominated by the SMBH and stellar masses, ignoring the contributions of other potential mass components such as gas and dark matter. This assumption is justified as the total molecular gas mass of NGC~1574 is less than $5\%$ of the total SMBH and stellar masses enclosed within the extent of the gas disc (\citealp{Ruffa_2023}; much less within the SMBH SoI), while the dark matter mass is usually negligible at the centres of massive galaxies \citep[e.g.][]{Cappellari_2013, Zhu_2024}.

Our stellar mass distribution model is identical to that of \citet{Ruffa_2023}. A multi-Gaussian expansion (MGE; \citealt{EMB_1994}) was performed on a \textit{Hubble Space Telescope} (\textit{HST}) Wide-Field Camera 2 F606W-band image, using the Python version of the \texttt{mge\_fit\_sectors\_regularized} procedure\footnote{Version 5.0 from \url{https://pypi.org/project/mgefit/}} of \citet{Cappellari_2002a}, to model the projected stellar light distribution of NGC~1574. The parameters of the best-fitting two-dimensional (2D) Gaussians, spatially deconvolved from \textit{HST}'s point-spread function (computed using the \textsc{TinyTim} package\footnote{Version 7.5 from \url{https://github.com/spacetelescope/tinytim/releases/tag/7.5}}; \citealt{Krist_2011}), are listed in Table~C1 of \citet{Ruffa_2023}. In each forward model, the 2D Gaussians are analytically deprojected to a three-dimensional (3D) axisymmetric light distribution given the inclination $i$, a free parameter of the model. The 3D stellar light distribution is then multiplied by a spatially constant mass-to-light ratio ($M/L$), another free parameter of the model, to obtain the 3D stellar mass volume density. Here, we assume that the light in the F606W band is not contaminated by the galactic nucleus, as there is no sign of an AGN in this galaxy from any existing observation. Finally, we add a freely-varying point mass ($M_\mathrm{BH}$) at the centre to represent the SMBH, yet another free parameter of the model.

We assume that the gas particles are in circular motions when deriving the gas kinematics from the input mass distribution. This assumption is largely consistent with the observed gas kinematics of NGC~1574, despite the signs of mild non-circular motions discussed in Section~\ref{subsec: uncertainty}. We compute the circular velocity curve corresponding to the combined (stellar + SMBH) mass distribution using the \texttt{{mge\_vcirc}} procedure of the \textsc{Jeans Anisotropic Modelling} (\textsc{JAM}) package\footnote{Version 7.2 from \url{https://pypi.org/project/jampy/}} of \citet{Cappellari_2008}. 

Finally, we add a spatially-constant velocity dispersion ($\sigma_\mathrm{gas}$) to the resultant circular velocity curve, another free parameter of the model. We assume that the gas velocity dispersion does not vary spatially because, as shown in Figure~\ref{fig:moments}, potential variations (except the artificial central increase due to beam smearing) are much smaller than the channel width of $20$~\kms\,and the central velocity enhancement caused by the SMBH. Models with spatially variable velocity dispersion profiles thus do not fit the data better nor noticeably change the best-fitting SMBH mass presented in Section~\ref{subsec: results}.

\subsection{Position angle warp}

To explain the observed kinematic twist of the gas disc, we adopt a PA warp model in which the PA varies linearly with radius from $\mathrm{PA}_\mathrm{inner}$ at $R=0$ (the disc's centre) to $\mathrm{PA}_\mathrm{outer}$ at $R=1\farcs2$ (approximately the outer edge of the disc). This warp model is the same as the one adopted by \citet{Ruffa_2023}. Yet, the innermost PA that we can model will be closer to the true PA at $R=0$, thanks to the higher spatial resolution of our data. This should improve the precision and accuracy of our measurements.

Our final dynamical model thus has a total of five free parameters describing the molecular gas disc ($i$, $\mathrm{PA}_\mathrm{inner}$, $\mathrm{PA}_\mathrm{outer}$, $\sigma_\mathrm{gas}$ and $R_\mathrm{s}$), two parameters describing the mass distribution ($M_\mathrm{BH}$ and $M/L$) and four "nuisance" parameters (offsets of the kinematic centre from the observation's phase centre, systemic velocity of the galaxy and total intensity of the molecular gas disc).

\subsection{Results}
\label{subsec: results}

\begin{figure*}
    \centering
    \includegraphics[width=\linewidth]{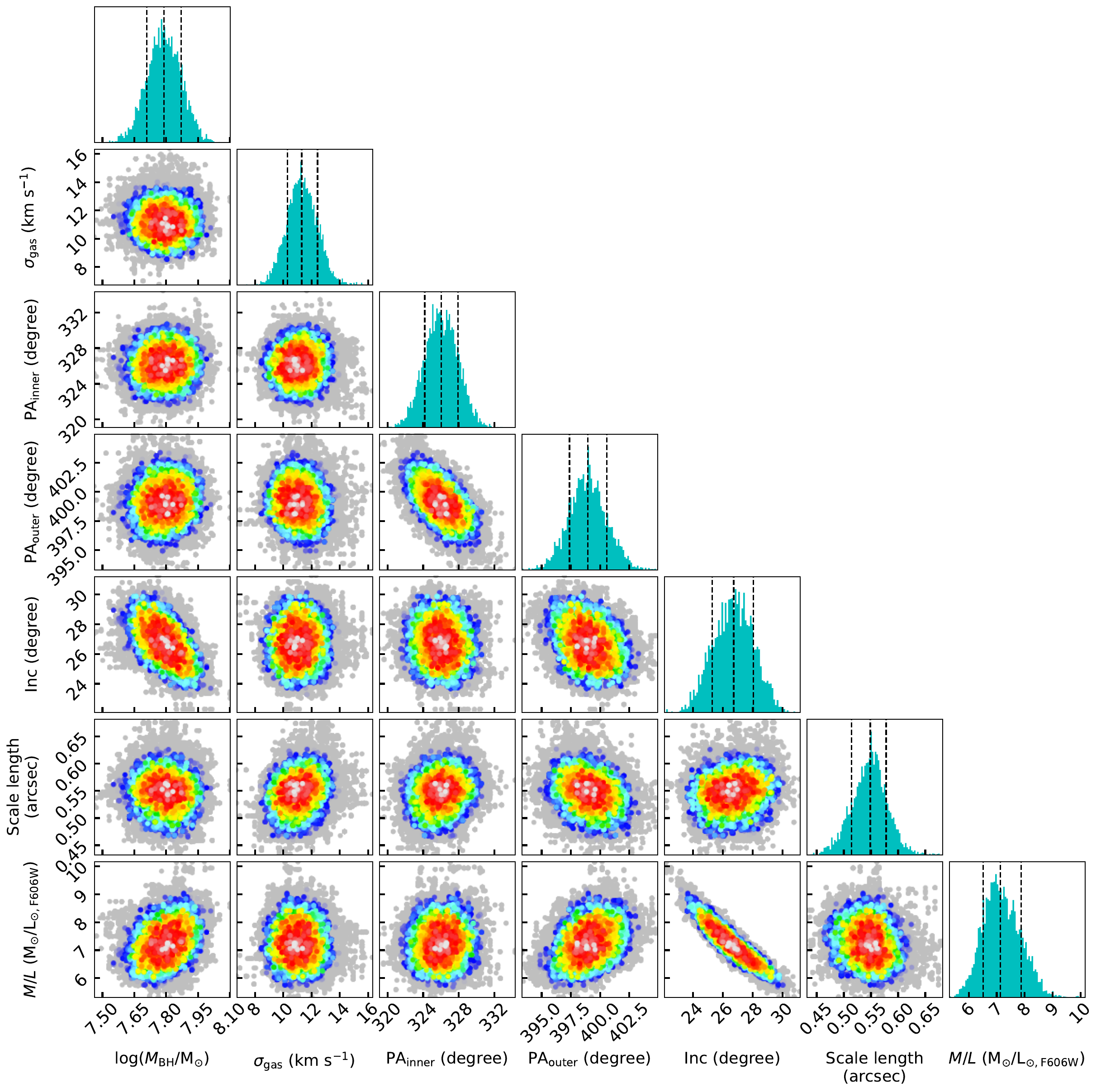}
    \caption{Corner plots showing the 2D joint posterior distributions between non-nuisance model parameters. The colours represent increasing confidence levels from $68.3\%$ (red, $1\sigma$) to $99.7\%$ (grey, $3\sigma$). The histograms show each parameter's 1D marginalised posterior distribution; the dashed lines indicate the medians and the $1\sigma$ confidence intervals.
    }
    \label{fig:corner}
\end{figure*}

We fit the dynamical model to the entire data cube (without applying any mask) using an MCMC chain of $2\times10^5$~steps. We discard the first $10^4$ steps as the burn-in to ensure the chain has converged, and confirm that the posterior distribution of each parameter is fully sampled after $2\times10^5$ steps. Table~\ref{tab:results} lists the best-fitting parameters and their $1\sigma$ ($68.3\%$ confidence level) and $3\sigma$ ($99.7\%$ confidence level) uncertainties. Figure \ref{fig:corner} shows each non-nuisance parameter's one-dimensional (1D) marginalised posterior distribution and the 2D joint posterior distributions between them. 

\begin{table*}
    \centering
    \caption{Best-fitting model parameters and associated uncertainties.}
    \begin{tabular}{l@{\hskip 1.7cm}c@{\hskip 1.7cm}c@{\hskip 1.7cm}c@{\hskip 1.7cm}c} \hline
    Parameter & Search~~range & Best fit & $1\sigma$ uncertainty & $3\sigma$ uncertainty \\ \hline
    \multicolumn{5}{l}{\textbf{Mass model}} \\ $\log\left(M_\mathrm{BH}/\mathrm{M}_\odot\right)$ & $6$ -- $10$ & $7.79$ & $\pm\,0.08$ & $-0.24$, $+0.22$ \\
    Stellar $M/L$ ($\mathrm{M}_\odot/\mathrm{L}_{\odot\mathrm{,F606W}}$) & $1$ -- $15$ & $7.1$ & $-0.6$, $+0.7$ & $-1.6$, $+2.4$ \\ \hline
    \multicolumn{5}{l}{\textbf{Molecular gas disc}} \\
    Inner position angle (degree) & $300$ -- $380$ & $326.0$ & $-1.8$, $+1.9$ & $-5.2$, $+5.5$ \\
    Outer position angle (degree) & $0$ -- $70$ & $39.0$ & $\pm\,1.6$ & $-5.0$, $+5.3$ \\
    Inclination (degree) & $10$ -- $40$ & $26.7$ & $-1.4$, $+1.3$ & $-3.7$, $+3.5$ \\
    Velocity dispersion (\kms) & $5$ -- $40$ & $11.3$ & $-1.0$, $+1.1$ & $-3.0$, $+3.9$ \\ 
    Scale length (arcsec) & $0.1$ -- $1.2$ & $0.55$ & $\pm\,0.03$ & $-0.10$, $+0.12$ \\ \hline
    \multicolumn{5}{l}{\textbf{Nuisance parameters}} \\
    Disc integrated intensity (Jy~\kms) & $5$ -- $30$ & $13.1$ & $\pm\,0.8$ & $-2.5$, $+2.4$ \\
    Kinematic centre $X$ offset (arcsec) & $-1.5$ -- $1.5$ & $0.00$ & $\pm\,0.01$ & $\pm\,0.02$ \\
    Kinematic centre $Y$ offset (arcsec) & $-1.5$ -- $1.5$ & $0.02$ & $\pm\,0.01$ & $\pm\,0.02$ \\
    Systemic velocity (\kms) & $950$ -- $1100$ & $1028.7$ & $\pm\,1.2$ & $-3.8$, $+3.5$ \\ \hline
    \end{tabular}\\
    {\textsl{Notes.} The $X$ and $Y$ offsets are measured relative to the phase centre of the data cube, $\mathrm{RA}=04^\mathrm{h}21^\mathrm{m}58\fs75$, $\mathrm{Dec.}=-56\degree58\arcmin28\farcs43$ (J2000.0)}.
    \label{tab:results}
\end{table*}

We obtain a best-fitting SMBH mass $M_\mathrm{BH}=(6.2\pm1.2)\times10^7$~M$_\odot$ and a F606W-band $M/L$ of ${7.13}_{-0.62}^{+0.75}$~$\mathrm{M}_\odot/\mathrm{L}_{\odot, \mathrm{F606W}}$. The PA changes linearly from $\mathrm{PA}_\mathrm{inner}=326\fdg0_{-1\fdg8}^{+1\fdg9}$ at $R=0$ to $\mathrm{PA}_\mathrm{outer}=39\fdg0 \pm 1\fdg6$ at $R=1\farcs2$ (after going through $360\degree$). Our SMBH mass is $1.7\sigma$ smaller than but is statistically consistent with the SMBH mass of $(1.0\pm0.2)\times10^8$~M$_\odot$ derived from the intermediate-resolution data of \citet{Ruffa_2023}, that did not spatially resolve the SoI. The $M/L$ is high but consistent with the $B-R$ colour of the galaxy $B-R=1.47\pm0.23$ \citep{Gao_2022}. According to the $M/L$-colour relation of \citet{Bell_2003} from stellar population synthesis models, the relation between $B-R$ colour and $V$-band $M/L$ is $\log[(M/L_V)/(\mathrm{M}_\odot/\mathrm{L}_{\odot,\mathrm{V}})] = 0.816(B-R)-0.633$. Thus, the expected $V$-band $M/L$ is $M/L_V = 3.7_{-1.3}^{+2.0}\,_{-0.5}^{+1.4}$\,$\mathrm{M}_\odot/\mathrm{L}_{\odot,\mathrm{V}}$, where the first uncertainty is statistical and the second uncertainty results from the $\approx0.1$~dex scatter of the relation. Our dynamically derived $M/L_\mathrm{F606W}$ is thus consistent with the range expected from the colour of the stellar population. It also agrees nicely with $M/L_\mathrm{F606W}=7.33_{-0.83}^{+0.92}$ \,$\mathrm{M}_\odot/\mathrm{L}_{\odot,\mathrm{F606W}}$ from \citet{Ruffa_2023}. However, the innermost PA is smaller than the $342\degree\pm0\fdg02$ of \citet{Ruffa_2023} by $8\sigma$, implying that the previous model underestimated the amplitude of the PA warp in the central region (as expected, given that it was spatially unresolved in their data). Nevertheless, the PA warp within the previously unresolved central $0\farcs2\times0\farcs2$ region is still part of the large-scale linear PA warp. A two-component PA warp model that allows both the PA and its radial gradient to change abruptly at $R=0\farcs2$ does not result in significantly different inner and outer PAs nor improves the Bayesian information criterion (BIC; $\mathrm{BIC}\equiv k\ln N+\chi^2$, where $k$ is the number of model parameters). The rest of the model parameters all agree within the uncertainties with those of \citet{Ruffa_2023}.

\begin{figure*}
    \centering
    \subfloat[][\phantom{xxxxxxxxxxx}]{
    \includegraphics[width=0.48\linewidth]{Figures/NGC1574_moment1.pdf}
    }
    \\
    \subfloat[][\phantom{xxxxxxxxxxx}]{
    \includegraphics[width=0.48\linewidth]{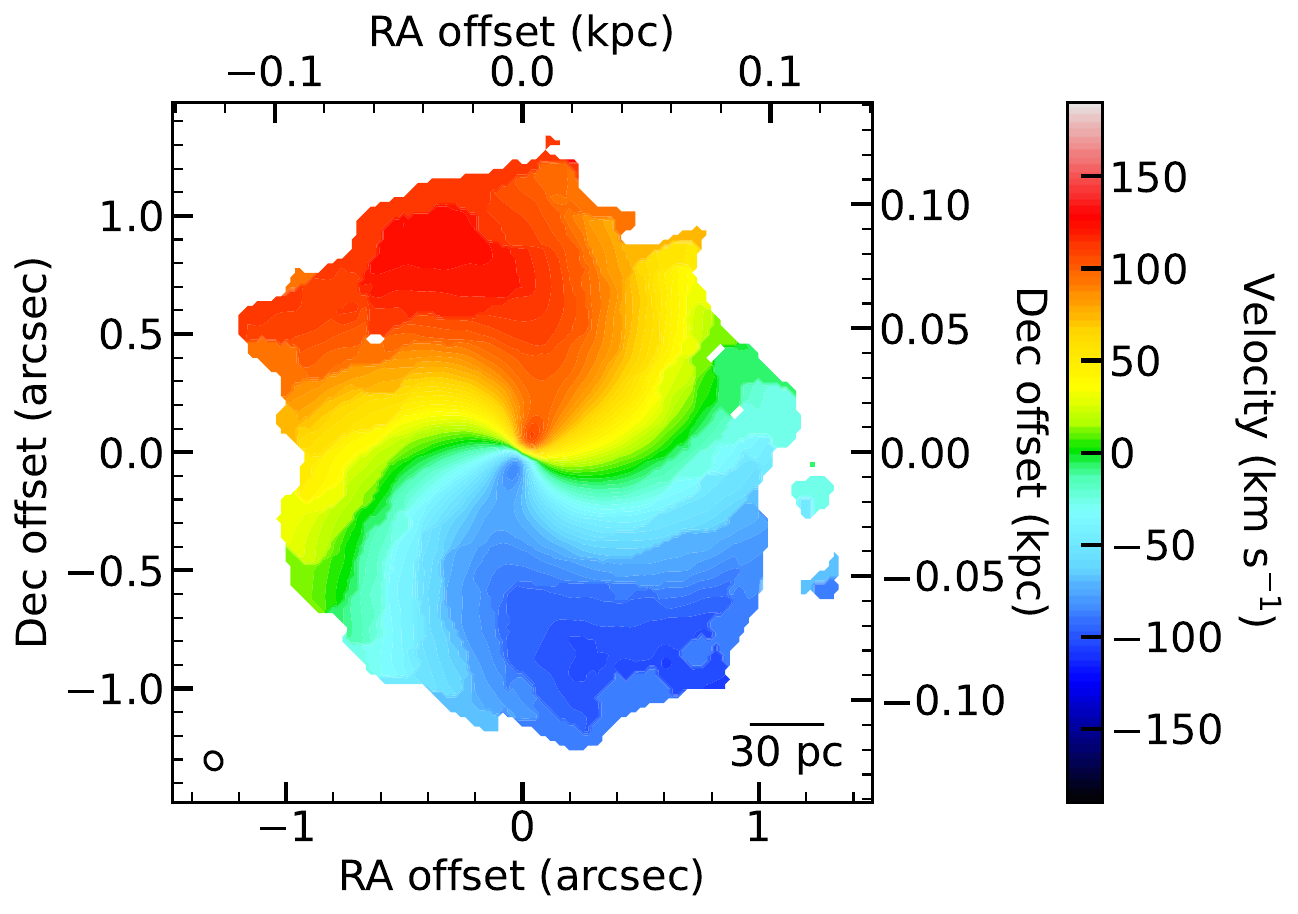}
    }
    \subfloat[][\phantom{xxxxxxxxxxx}]{
    \includegraphics[width=0.48\linewidth]{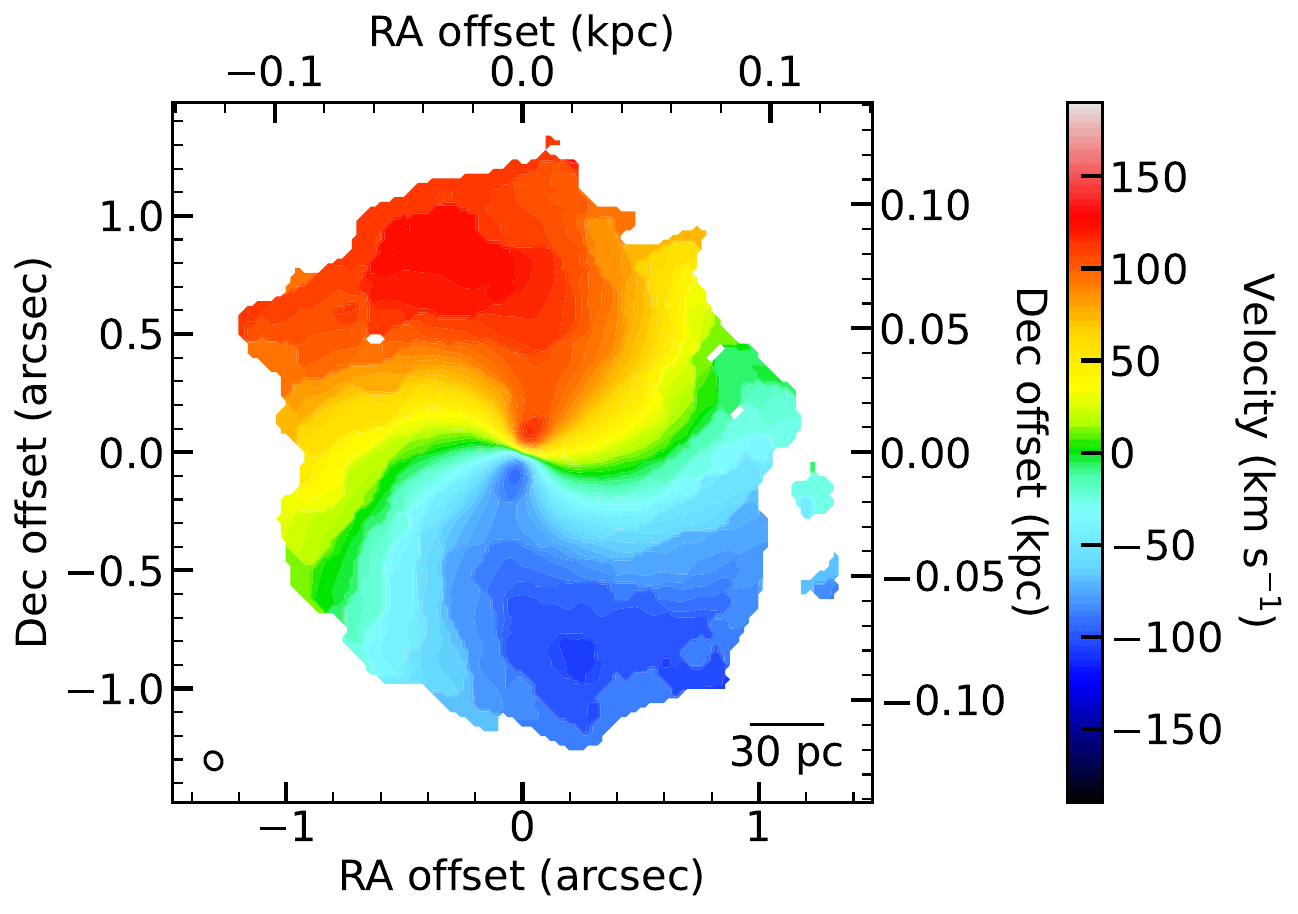}
    } \\
    \subfloat[][\phantom{xxxxxxxxxx}]{
    \includegraphics[width=0.48\linewidth]{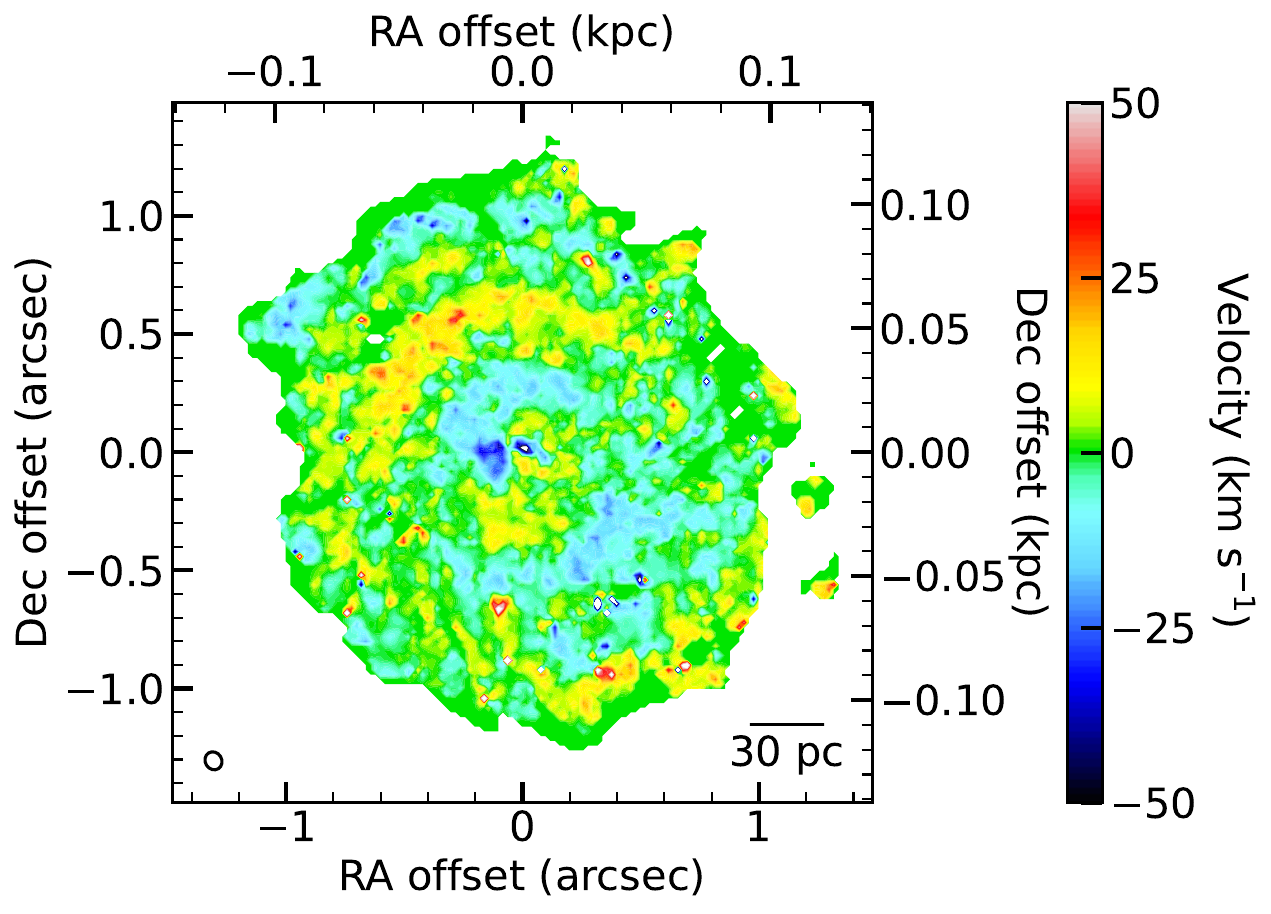}
    } 
    \subfloat[][\phantom{xxxxxxxxxx}]{
    \includegraphics[width=0.48\linewidth]{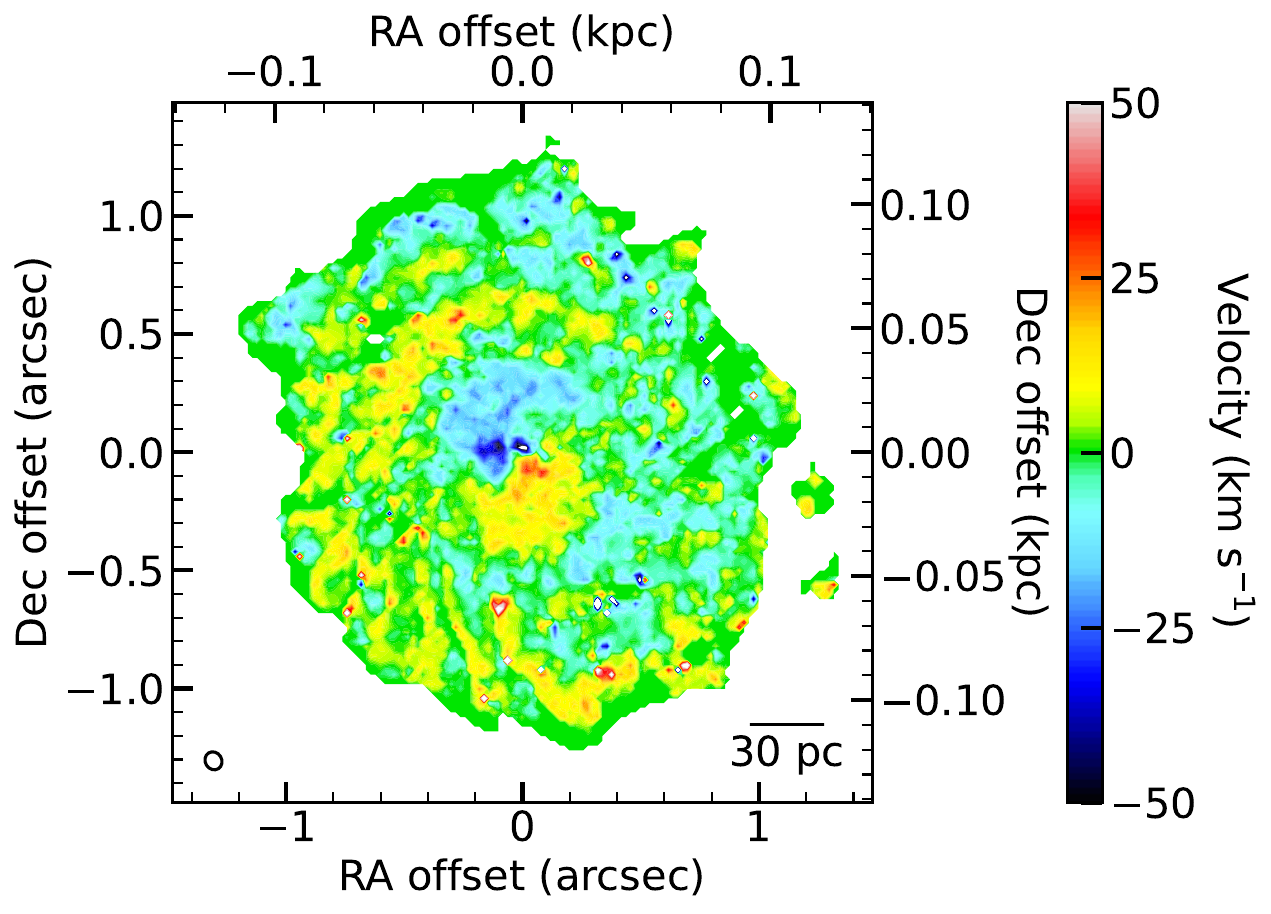}
    } \\
    \caption{
    (a) First-moment (intensity-weighted mean line-of-sight velocity) map of the NGC~1574 data cube. (b) -- (c) First-moment maps of the best-fitting dynamical model from (b) this work and (c) the best-fitting model of \citeauthor{Ruffa_2023} (\citeyear{Ruffa_2023}), obtained by fitting intermediate spatial resolution observations. (d) -- (e) Corresponding first-moment residual maps ($\mathrm{data}-\mathrm{model}$). The synthesised beam ($0\farcs078\times0\farcs069$) is shown as a black open ellipse in the bottom-left corner of each panel, while a $30$~pc scale bar is shown in the bottom-right corner of each panel. Positions are measured relative to the best-fitting kinematic centre; velocities relative to the best-fitting systemic velocity. The model derived using only intermediate-resolution observations slightly underestimates the amplitude of the PA warp in the central $0\farcs2\times0\farcs2$ region, leading to larger velocity residuals.}
    \label{fig:velo_resi}
\end{figure*}

Figure~\ref{fig:velo_resi} compares the (luminosity-weighted mean line-of-sight) velocity map derived from the high-resolution ALMA data of NGC~1574 (panel a) with the equivalent velocity map of the best-fitting model from this work (panel b) and that of the best-fitting model of \citeauthor{Ruffa_2023} (\citeyear{Ruffa_2023}; panel c). Panels (d) and (e) show the corresponding velocity residual ($\mathrm{data}-\mathrm{model}$) maps of the two models. The model derived using only the intermediate-resolution data slightly underestimates the amplitude of the PA warp in the previously unresolved central $0\farcs2\times0\farcs2$ region, leading to larger velocity residuals.

\begin{figure*}
    \captionsetup[subfigure]{labelformat=empty}
    \centering
    \subfloat[]{
    \includegraphics[width=0.3335\linewidth]{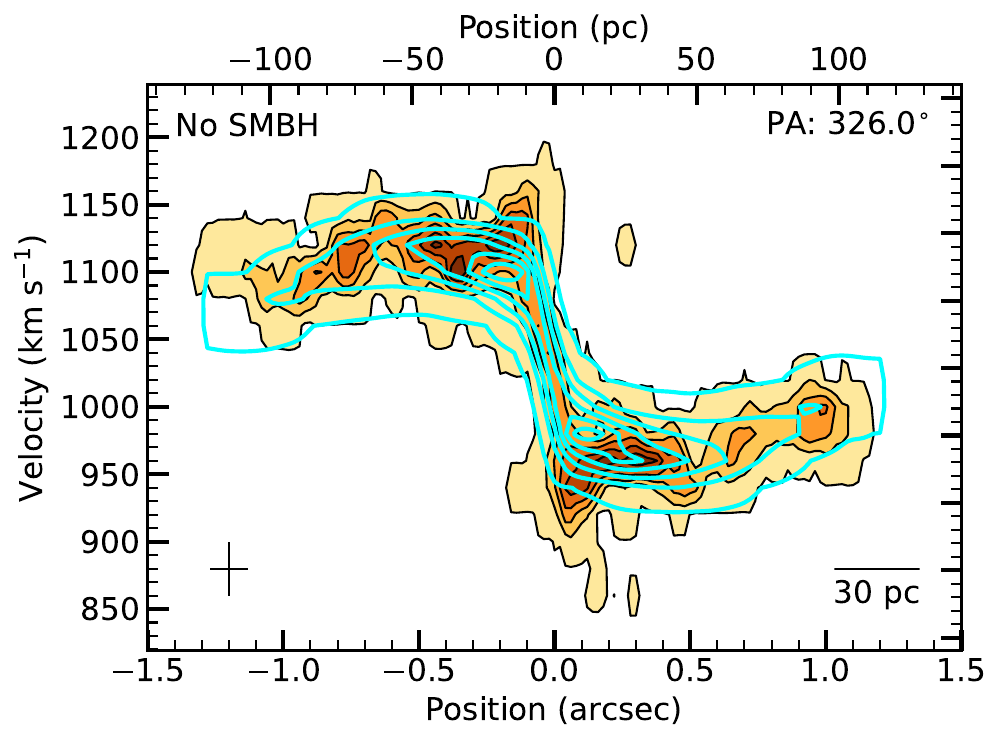}
    }
    \subfloat[]{
    \includegraphics[width=0.302\linewidth]{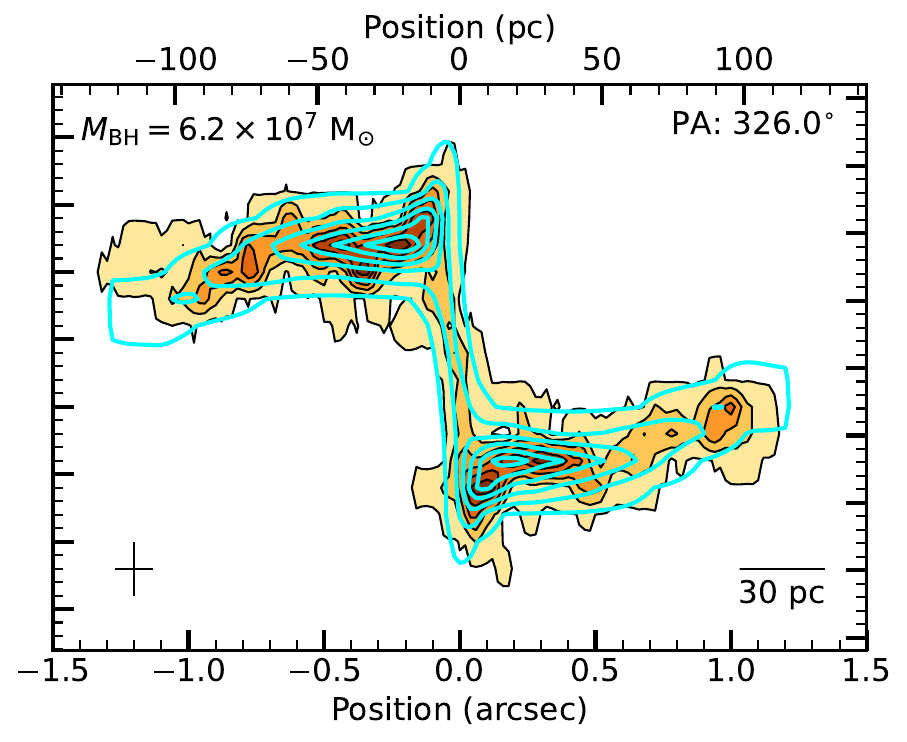}
    }
    \subfloat[]{
    \includegraphics[width=0.3395\linewidth]{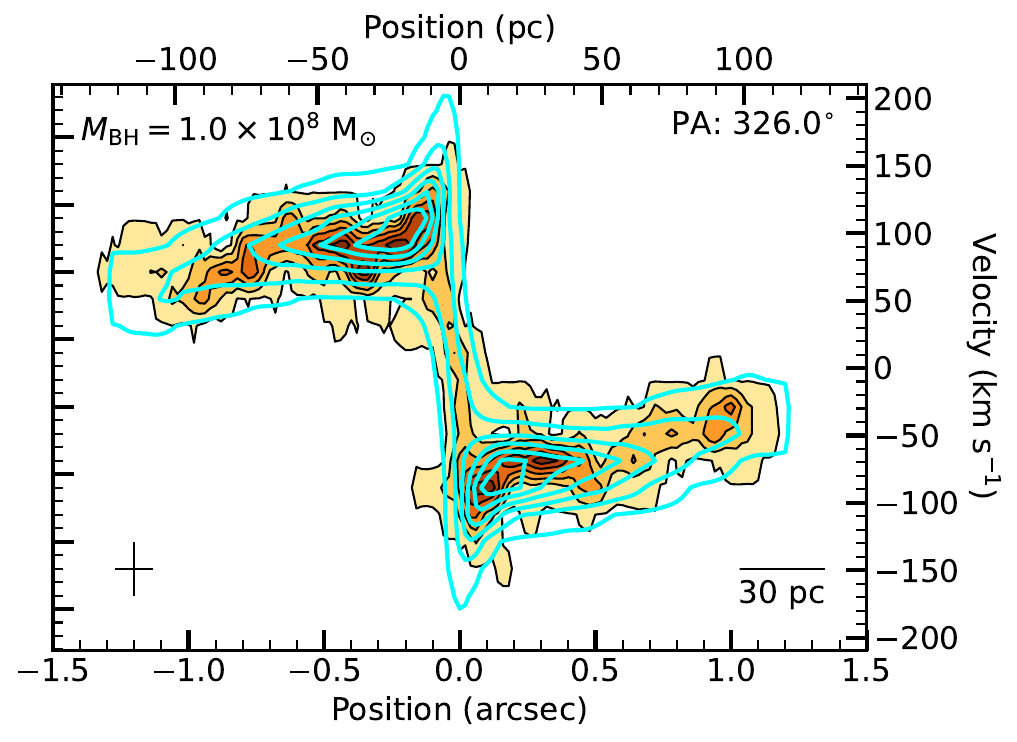}
    }
    \caption{
    Observed kinematic major-axis position-velocity diagram of NGC~1574 (orange scale with black contours), overlaid with the PVDs of different models (cyan contours): no SMBH (left), best-fitting model from this work (centre) and best-fitting model of \citet{Ruffa_2023} derived using only intermediate spatial resolution observations that do not spatially resolve the SMBH SoI (right). Positions are measured relative to the best-fitting kinematic centre; velocities relative to the best-fitting systemic velocity along the right axis. The cross in the bottom-left corner of each panel shows the synthesised beam FWHM along the kinematic major axis and the channel width. A scale bar is shown in the bottom-right corner of each panel. Our best-fitting model reproduces the Keplerian velocity increase toward the galaxy centre nicely, but the best-fitting model of \citet{Ruffa_2023} overshoots the velocity rise. This demonstrates that high-resolution observations that spatially resolve the SoI are necessary for accurate SMBH mass measurements.}
    \label{fig:pvd+model}
\end{figure*}

Figure \ref{fig:pvd+model} then compares the kinematic major-axis PVD of the high-resolution data with those of three different models: a model with no SMBH (left), the best-fitting model from this work (centre) and the best-fitting model of \citet{Ruffa_2023} with a slightly more massive SMBH. The model with no SMBH fails to reproduce the central velocity rise, while the best-fitting model derived using only the intermediate-resolution data overshoots the velocity increase due to its slightly larger SMBH mass. By contrast, the best-fitting model from this work accurately reproduces the PVD at most positions and velocities. The slight difference between the width of the data PVD and that of the model PVD near a major-axis position of $0\farcs1$ is likely due to small non-circular motions, which are also noticeable in the velocity residual map (panel d of Figure~\ref{fig:velo_resi}; discussed further in Section~\ref{subsec: uncertainty}).

We finally compare the models' reduced $\chi^2$ ($\chi_\mathrm{r}^2$) statistics measured using the high-resolution data. Our new best-fitting model has $\chi_\mathrm{r}^2=2.14$, while the previous model of \citet{Ruffa_2023} has $\chi_\mathrm{r}^2=2.37$ (with the same free parameters). This new measurement of NGC~1574 thus demonstrates the importance of high-resolution observations, that spatially-resolve the SMBHs' SoI, for accurate SMBH mass measurements and dynamical models.

\section{Discussion}
\label{sec: discuss}

\subsection{Main uncertainties} \label{subsec: uncertainty}

The statistical uncertainty of the SMBH mass in NGC~1574 is substantially increased by the uncertainty of the inclination. This is because $M_\mathrm{BH}$ is proportional to the square of the deprojected rotational velocity $v_\mathrm{rot}$, $M_\mathrm{BH}\propto v_\mathrm{rot}^2$, and $v_\mathrm{rot}$ is related to the observed line-of-sight velocity $v_\mathrm{obs}$ by $v_\mathrm{rot}=v_\mathrm{obs}/\sin{i}$, so that $M_\mathrm{BH}\propto \sin^{-2}i$. This implies that the SMBH mass uncertainty is largest when the galaxy is nearly face-on, as is the case for NGC~1574 ($i\approx27\degree$). The SMBH mass uncertainty would shrink from $19\%$ to $16\%$ with no inclination uncertainty. 

Contrary to many other measurements, the SMBH mass of NGC~1574 is not anti-correlated with its stellar $M/L$, as our high spatial resolution observations spatially resolve the SMBH SoI and trace the inner Keplerian part of the rotation curve, which is largely dominated by the SMBH and is relatively independent of the stars. This breaks the mass degeneracy. The positive correlation between SMBH mass and stellar $M/L$ is then a by-product of the SMBH mass -- inclination anti-correlation and the $M/L$ -- inclination anti-correlation, as a lower inclination implies higher deprojected velocities and thus both a higher SMBH mass and a higher stellar mass \citep{Smith_2019}.

A potential source of systematic uncertainty in our measurement is the assumption that the gas particles move on circular orbits. Indeed, the residual (luminosity-weighted mean line-of-sight) velocity map of our model (panel d of Figure~\ref{fig:velo_resi}) shows spiral arm-like residuals up to $\pm 55$~\kms\ (after deprojection), suggesting mild non-circular motions and recovering the finding of \citet{Ruffa_2023}. This could result from the non-axisymmetric potential induced by the nuclear stellar bar of NGC~1574, potentially producing shocks along the bar edges that would lead to gas inflows/outflows down to the scale of the circumnuclear molecular gas disc observed. However, the non-circular motions are too small compared to the circular velocities detected within the SMBH SoI (maximum $\approx380$~\kms\ after deprojection) to substantially affect the SMBH mass measurement.

Another potential source of systematic uncertainty in our SMBH mass measurements is the effect of dust extinction on the MGE fit and associated stellar mass distribution \citep[e.g.][]{Boizelle_2021, Cohn_2024}, but the \textit{HST} image of NGC~1574 does not show any obvious sign of dust obscuration.

As is customary for dynamical SMBH mass measurements, we choose not to include in the SMBH mass uncertainty budget the $10\%$ uncertainty introduced by the distance, which is directly proportional to $\mbh$. Rescaling our results to a different distance therefore does not require redoing any dynamical modelling.

\subsection{Resolving the SMBH sphere of influence}
\label{subsec:SoI}

An empirical approximation for the radius of a SMBH's SoI is $R_\mathrm{SoI}\equiv GM_\mathrm{BH}/\sigma_\mathrm{e}^2$. Given our updated SMBH mass (see Section~\ref{subsec: results}) and $\sigma_\mathrm{e}=216\pm16$~\kms\ \citep{Bernardi_2002}, $R_\mathrm{SoI}$ is thus $5.7$~pc or $\approx0\farcs059$. The angular resolution of our observation, $0\farcs074$, is thus slightly worse than $R_\mathrm{SoI}$. However, $R_\mathrm{SoI}$ is only a proxy for the more formal and physically-motivated definition of the SoI, that is the equality radius $R_\mathrm{eq}$ at which the enclosed stellar mass equals the SMBH mass:
\begin{equation}
    M_*(R=R_\mathrm{eq})=M_\mathrm{BH}\,\,.
\end{equation}

\begin{figure}
    \centering
    \includegraphics[width=\linewidth]{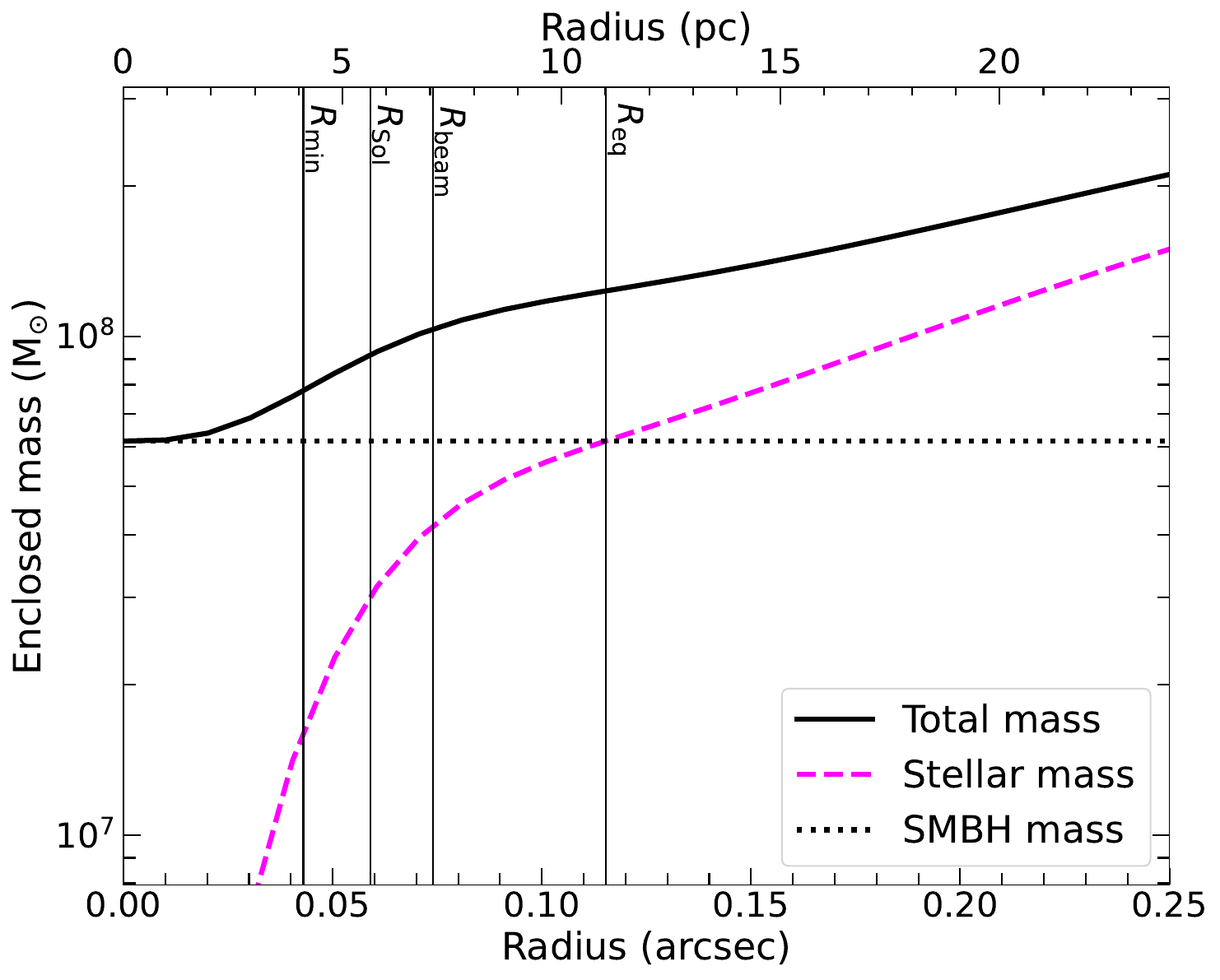}
    \caption{Cumulative mass distribution of NGC~1574, showing the relative contributions of stars (magenta dashed line) and the SMBH (black dotted line) to the total enclosed mass (black solid line). The four vertical black lines show the radius of the innermost detected kinematic tracer $R_\mathrm{min}$, the traditionally-defined radius of the SMBH SoI $R_\mathrm{SoI}$, the mean spatial extent of the synthesised beam $R_\mathrm{beam}$ and the radius at which the enclosed stellar mass equals the SMBH mass $R_\mathrm{eq}$.
    }
    \label{fig:mass_encl}
\end{figure}

To compute $R_\mathrm{eq}$, we derive the cumulative mass distribution of the stars using the MGE components of \citet{Ruffa_2023}, our best-fitting stellar $M/L$ and the \texttt{{mge\_radial\_mass}} procedure of the \textsc{JAM} package. We then calculate the radius where the cumulative stellar mass equals the SMBH mass, yielding $R_\mathrm{eq}=0\farcs12\pm0\farcs03$ ($11.0 \pm 2.6$~pc), where the uncertainty arises from the uncertainties of the stellar $M/L$ and SMBH mass. Figure~\ref{fig:mass_encl} compares the relative contributions of the stars and the SMBH to the total cumulative mass distribution of NGC~1574, and shows $R_\mathrm{eq}$ as a vertical black line (as well as $R_\mathrm{SoI}$ and the spatial extent of the synthesised beam, $R_\mathrm{beam}$). The high-resolution data presented in this paper spatially resolve $R_\mathrm{eq}$ by a factor of $R_\mathrm{eq}/R_\mathrm{beam}\approx1.6$, and are thus the first to fully spatially resolve the SMBH SoI of NGC~1574. For this reason, our SMBH mass measurement is more accurate than that reported by \citet{Ruffa_2023} using observations with an angular resolution of $0\farcs17$ (thus larger than both $R_\mathrm{SoI}$ and $R_\mathrm{eq}$). 

In practice, the radius of the innermost kinematic tracer that a molecular gas observation can detect is usually slightly smaller than $R_\mathrm{beam}$, as positions can be measured more precisely than the FWHM of the synthesised beam \citep{Zhang_2024}. We thus estimate the radius of the innermost detected kinematic tracer ($R_\mathrm{min}$) of our high-resolution observations from the PVD (Figure~\ref{fig:pvd+model}), following the procedure outlined by \citet{Zhang_2024}. This yields $R_\mathrm{min}\approx0\farcs043$ or 4.1~pc (shown by another vertical black line in Figure~\ref{fig:mass_encl}), smaller than both $R_\mathrm{SoI}$ ($R_\mathrm{min}/R_\mathrm{SoI}=0.72$) and $R_\mathrm{eq}$ ($R_\mathrm{min}/R_\mathrm{eq}=0.37$) and well within the region where the SMBH dominates over the stars. As a result, our observations probe deep into the region where the gravitational influence of the SMBH dominates the gas kinematics and detect a prominent Keplerian velocity increase, even though the synthesised beam only resolves $R_\mathrm{eq}$ by a factor of $\approx2$.

Although our high-resolution measurement does not formally provide a better statistical uncertainty ($\approx20\%$) than that of the previous intermediate-resolution measurement, this is partly because the previous measurement predicted a larger SMBH mass and was thus overconfident in the significance of the SMBH detection, resulting in an overconfident uncertainty. Other attempts to improve molecular-gas-dynamical SMBH mass measurements using observations that highly resolve the SoI (e.g.\,\citealt{Ruffa_2023} for NGC~4261 and \citealt{Zhang_2025} for NGC~383) all led to statistical uncertainties substantially better than those of lower-resolution measurements. These high-resolution measurements also revealed that the lower-resolution measurements tend to slightly overestimate the SMBH masses, implying that measurements that marginally resolve the SoI might be systematically inaccurate. Hence, it is also crucial to obtain precise and accurate SMBH mass measurements using observations that highly resolve the SoI to identify whether the observed scatters of the SMBH--galaxy scaling relations represent intrinsic scatters caused by variations of galactic properties \citep[e.g. morphology or mass;][]{McConnell_2013} or are dominated by measurement inaccuracies. Additionally, high-resolution observations of molecular gas around the SMBH can probe the dynamics within the circumnuclear disk down to a few times $10^4R_\mathrm{Sch}$, revealing physical processes directly related to the fuelling and feedback of SMBHs, such as radial motions and/or alignment between molecular gas disk and radio jet \citep{Zhang_2025}.

\subsection{Comparison with SMBH -- galaxy property scaling relations}

\begin{figure}
    \centering
    \includegraphics[width=\linewidth]{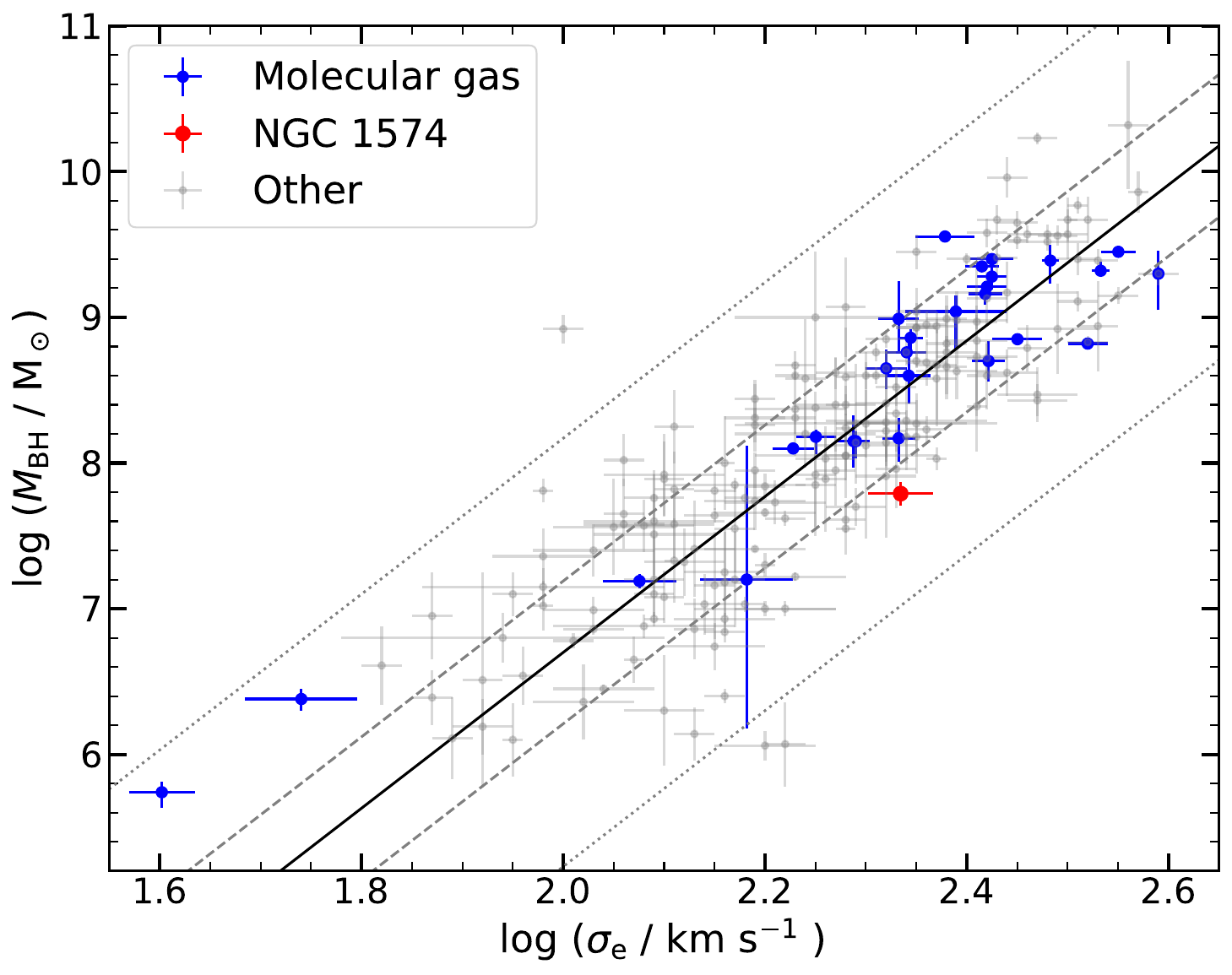}
    \caption{SMBH mass ($M_\mathrm{BH}$) and effective stellar velocity dispersion ($\sigma_\mathrm{e}$) of NGC~1574 (red data point), compared to those of other galaxies with SMBH mass measurements derived using molecular gas kinematics (blue data points) and other methods (grey data points). The best-fitting $M_\mathrm{BH}$ -- $\sigma_\mathrm{e}$ relation of \citet{Bosch_2016} is shown as a black solid line, with one and three times the observed scatter around the relation shown with dashed and dotted lines, respectively. Our SMBH mass measurement places NGC~1574 slightly below the relation's $1\sigma$ scatter.
    }
    \label{fig:M-sig}
\end{figure}

To place our SMBH mass measurement in the context of SMBH -- galaxy property scaling relations, we compare in Figure~\ref{fig:M-sig} the SMBH mass and effective stellar velocity dispersion ($\sigma_\mathrm{e}=216\pm16$~\kms; \citealt{Bernardi_2002}) of NGC 1574 (red data point) with the best-fitting $M_\mathrm{BH}$ -- $\sigma_\mathrm{e}$ relation of \citeauthor{Bosch_2016} (\citeyear{Bosch_2016}; black solid line). The dashed and the dotted lines show one and three times the observed scatter of the relation, respectively. We also show other galaxies with SMBH masses derived using molecular gas kinematics (blue data points; compiled by \citealt{Zhang_2024}) and other methods (grey data points; compiled by \citealt{Bosch_2016}). Our updated SMBH mass places NGC~1574 slightly below the observed $1\sigma$ scatter of the relation, indicating that the galaxy has one of the smallest SMBH masses among galaxies with $\sigma_\mathrm{e}\geq200$~\kms. This does not make NGC~1574 a significant outlier of the $M_\mathrm{BH}$ -- $\sigma_\mathrm{e}$ relation, as it is well within the observed $3\sigma$ scatter. However, the intrinsic scatter might be substantially smaller than the observed scatter and needs to be constrained by more precise and accurate SMBH mass measurements.

\begin{figure}
    \centering
    \includegraphics[width=\linewidth]{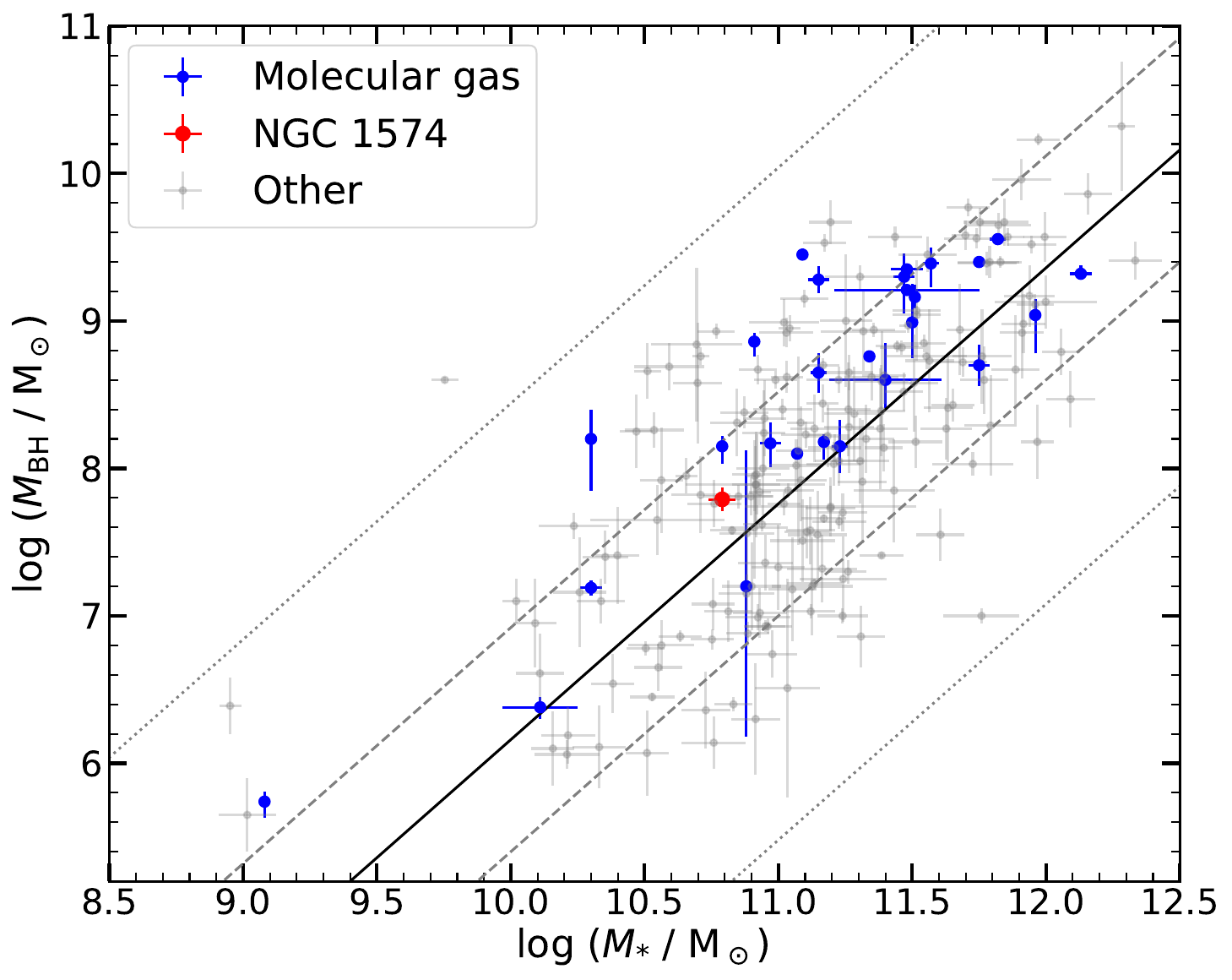}
    \caption{As Figure~\ref{fig:M-sig}, but for the SMBH masses ($M_\mathrm{BH}$) and total stellar masses ($M_\mathrm{*}$). Our SMBH mass measurement places NGC~1574 within the $1\sigma$ scatter of the relation.
    }
    \label{fig:BH-star}
\end{figure}

Figure~\ref{fig:BH-star} similarly compares the SMBH mass and total stellar mass $M_\mathrm{*}$ of NGC~1574 with those of other galaxies with SMBH mass measurements and with the best-fitting $M_\mathrm{BH}$ -- $M_\mathrm{*}$ relation of \citet{Bosch_2016}. The stellar masses of galaxies whose SMBHs were measured using the molecular gas method were derived from the published MGE models if available. Our dynamically derived SMBH mass of NGC~1574 is well within the observed $1\sigma$ scatter of the $M_\mathrm{BH}$ -- $M_\mathrm{*}$ relation. In other words, the SMBH of NGC~1574 is not less massive than that expected from the $M_\mathrm{BH}$ -- $M_\mathrm{*}$ relation. 
The slight offset of NGC~1574 from the mean $M_\mathrm{BH}$ -- $\sigma_\mathrm{e}$ relation is thus more likely the result of an overly large $\sigma_\mathrm{e}$ rather than a truly small $M_\mathrm{BH}$. This could be caused by motions induced by the stellar bar of the galaxy.

\section{Conclusions}
\label{sec: conclude}

We presented a measurement of the SMBH mass of NGC~1574, a lenticular galaxy hosting a nuclear stellar bar. ALMA observations of the $^{12}$CO(2–1) emission line with synthesised beam FWHM of $0\farcs078\times0\farcs070$ ($\approx7.5\times6.7$~pc$^2$) reveal a prominent velocity increase towards the centre of the molecular gas disc, a kinematic twist associated with a PA warp throughout the disc, and evidence of mild ($<15\%$ of the typical circular velocities) non-circular motions at the very centre, potentially induced by the nuclear bar. Our forward dynamical modelling of the data cube yields a SMBH mass $M_\mathrm{BH}=(6.2\pm1.2)\times10^7$~M$_\odot$, slightly smaller than but statistically consistent with the previous SMBH mass of $(1.0\pm0.2)\times10^8$~$\msun$ derived using only intermediate-resolution data that do not spatially resolve the SMBH SoI ($R_\mathrm{SoI}$ or $R_\mathrm{eq}$; \citealt{Ruffa_2023}). Our new SMBH mass places NGC~1574 slightly below the $1\sigma$ scatter of the $M_\mathrm{BH}$ -- $\sigma_\mathrm{e}$ relation of \citet{Bosch_2016}. The PA warp is most consistent with a linear radial variation from $\mathrm{PA}_\mathrm{inner}=326\fdg0_{-1\fdg8}^{+1\fdg9}$ at $R=0$ to $\mathrm{PA}_\mathrm{outer}=39\fdg0 \pm 1\fdg6$ at $R=1\farcs2$, suggesting that the PA warp within the central $0\farcs2\times0\farcs2$ region is larger than previously thought. The F606W-band $M/L$ is ${6.88}_{-0.49}^{+0.69}$~$\mathrm{M}_\odot/\mathrm{L}_{\odot, \mathrm{F606W}}$, agreeing with the result of \citet{Ruffa_2023} within $1\sigma$.

Our new high spatial resolution measurement of the SMBH mass of NGC~1574 does not improve the precision of the SMBH mass ($\approx20\%$), but it considerably improves its accuracy and the PA warp model, as our observations are the first to spatially resolve the SMBH SoI of NGC~1574 in terms of both the synthesised beam size $R_\mathrm{beam}$ ($R_\mathrm{beam}/R_\mathrm{eq}=0.64$) and the radius of the innermost detected kinematic tracer $R_\mathrm{min}$ ($R_\mathrm{min}/R_\mathrm{eq}=0.37$). Our study thus again emphasises the importance of high-resolution observations, that fully spatially resolve the gravitational spheres of influence of the SMBHs, for accurate $\mbh$ measurements and gas dynamical models.

\section*{Acknowledgements}

HZ acknowledges support from a Science and Technology Facilities Council (STFC) DPhil studentship under grant ST/X508664/1 and the Balliol College J.\ T.\ Hamilton Scholarship in physics.
MB was supported by STFC consolidated grant ‘Astrophysics at Oxford’ ST/K00106X/1 and ST/W000903/1. IR and TAD acknowledge support from STFC consolidated grant ST/S00033X/1.
This paper makes use of the following ALMA data: ADS/JAO.ALMA\#2015.1.00419.S, ADS/JAO.ALMA\#2016.2.00053.S and ADS/JAO.ALMA\#2022.1.01122.S. 
ALMA is a partnership of ESO (representing its member states), NSF (USA) and NINS (Japan), together with NRC (Canada), MOST and ASIAA (Taiwan), and KASI (Republic of Korea), in cooperation with the Republic of Chile. The Joint ALMA Observatory is operated by ESO, AUI/NRAO and NAOJ. This research made use of the NASA/IPAC Extragalactic Database (NED), which is operated by the Jet Propulsion Laboratory, California Institute of Technology, under contract with the National Aeronautics and Space Administration.


\section*{Data Availability}

The raw data used in this study are publicly available via the ALMA Science Archive (\url{https://almascience.eso.org/aq/}) and the Hubble Science Archive (\url{https://hst.esac.esa.int/ehst/}). The calibrated data, final data products and original plots generated for this research will be shared upon reasonable requests to the
first author.


\bibliographystyle{mnras}
\bibliography{refs} 

\bsp	
\label{lastpage}
\end{document}